\documentclass[aps,prd,showpacs,showkeys,nofootinbib,preprintnumbers,notitlepage,superscriptaddress]{revtex4-1}
\usepackage{amsmath,amssymb,amsthm,bm,bbm,color,hyperref}
\usepackage{graphicx,subfigure,rotating,capt-of}

\begin{document} 

\title{Confining Bond Rearrangement in the Random Center Vortex Model}

\author{Derar Altarawneh}
\email{derar@nmsu.edu}
\affiliation{Department of Physics, New Mexico State University, PO Box 30001, Las Cruces, NM 88003-8001, USA}
\affiliation {Department of Applied Physics, Tafila Technical University, Tafila , 66110 , Jordan}
\author{Roman H\"ollwieser\footnote{Funded by an Erwin Schr\"odinger Fellowship of the Austrian Science Fund under Contract No. J3425-N27.}}
\email{hroman@kph.tuwien.ac.at}
\affiliation{Department of Physics, New Mexico State University, PO Box 30001, Las Cruces, NM 88003-8001, USA}
\affiliation{Institute of Atomic and Subatomic Physics, Nuclear Physics Dept.\\ Vienna University of Technology, Operngasse 9, 1040 Vienna, Austria}
\author{Michael Engelhardt}
\email{engel@nmsu.edu}
\affiliation{Department of Physics, New Mexico State University, PO Box 30001, Las Cruces, NM 88003-8001, USA}

\date{\today}

\begin{abstract}
We present static meson-meson and baryon--anti-baryon potentials in $Z(2)$
and $Z(3)$ random center vortex models for the infrared sector of Yang-Mills
theory, {\it i.e.}, hypercubic lattice models of random vortex world-surfaces.
In particular, we calculate multiple Polyakov loop correlators corresponding
to static meson-meson resp.~baryon--anti-baryon configurations in a center
vortex background and observe that their expectation values follow the minimal
area law, displaying bond rearrangement behavior, a characteristic expected
for the confining dynamics of the strong interaction. The static meson-meson
and baryon--anti-baryon potentials are compared with theoretical predictions
and lattice QCD simulations.
\end{abstract}



\maketitle

\tableofcontents

\newpage

\section{Introduction}

Center vortices~\cite{'tHooft:1977hy,Vinciarelli:1978kp,
Yoneya:1978dt,Cornwall:1979hz,Mack:1978rq,Nielsen:1979xu} are closed tubes of
quantized chromomagnetic flux which spontaneously condense in the vacuum, giving
rise to the non-perturbative phenomena which characterize the infrared sector of
strong interaction physics, namely, quark confinement, the spontaneous breaking
of chiral symmetry ($\chi$SB) and the axial $U_A(1)$ anomaly.

The vortex model of confinement has been buttressed by
a multitude of studies, in lattice Yang-Mills theory, see {\it
e.g.}~\cite{DelDebbio:1996mh,Langfeld:1997jx,DelDebbio:1997ke,DelDebbio:1998uu,
Kovacs:1998xm,Alexandrou:1999iy,Engelhardt:1999fd,Bertle:2001xd, Bertle:2002mm}, 
within the infrared effective model of center vortices under investigation here~
\cite{Engelhardt:1999wr,Engelhardt:2000wc,Engelhardt:2002qs,Engelhardt:2003wm,Quandt:2004gy,Engelhardt:2004qq,Engelhardt:2005qu,Engelhardt:2006ep,Engelhardt:2010ft} and another infrared model of random vortex lines in continuous 3D
space-time~\cite{Altarawneh:2014aa,Hollwieser:2014lxa} and also by theoretical
effective model calculations,
e.g.~\cite{Oxman:2012ej,Oxman:2014sea,Nejad:2014tja,Oxman:2015oba}.
The physical motivation for the vortex picture can be seen from the Wilson loop
criterion of confinement, {\it i.e.}, the expectation value of the Wilson loop
has to follow an area law in the confined phase. The vortex model
deduces the area law from independent vortex piercings of the Wilson loop, which
can be interpreted as crossings of the static electric flux tube and moving
closed magnetic flux. The area law follows from fluctuations in the number of
vortices piercing a Wilson loop and the deconfinement transition in the center
vortex picture takes the guise of a percolation transition~\cite{Engelhardt:1999fd,Engelhardt:1999wr}.

Numerical simulations indicate further that vortices are responsible for $\chi$SB
as well~\cite{deForcrand:1999ms,Alexandrou:1999vx,Engelhardt:1999xw,Engelhardt:2002qs,Leinweber:2006zq,
Bornyakov:2007fz,Jordan:2007ff,Hollwieser:2008tq,Bowman:2010zr,Hollwieser:2009wka,Hollwieser:2010mj,Engelhardt:2010ft,
Hollwieser:2011uj,O'Malley:2011zz,Hollwieser:2012kb,OMalley:2011aa,Schweigler:2012ae,Hollwieser:2013xja,Trewartha:2014ona,Faber2014lxa,Brambilla:2014jmp,Nejad:2015aia,Trewartha:2015nna,Trewartha:2015ida}.
Removal of center vortices restores chiral symmetry~\cite{deForcrand:1999ms}. 
A dense vortex vacuum gives rise to a finite density of Dirac operator
near-zero modes, see also~\cite{Hollwieser:2014mxa,Hollwieser:2015koa},
hence a finite chiral condensate via the Banks-Casher
relation~\cite{Banks:1979yr}, spontaneously breaking chiral symmetry.
More than one concrete mechanism may be at play in generating the
near-zero mode spectrum. On the one hand, center vortices give rise to lumps
of topological charge through (self-)intersection and writhe, as well
as through their color structure. These attract Dirac zero modes, which
can interact to generate the near-zero mode spectrum in a manner reminiscent
of the instanton liquid model, as studied in ~\cite{Hollwieser:2013xja},
where the close analogy between specific spherical vortex configurations
and instantons was noted. On the other hand, even in the absence of
would-be zero modes generated by topological charge lumps, the random
interactions of quarks with the vortex background may be strong enough
to smear the free quark dispersion relation such that the spectral density
of the Dirac operator becomes nonzero near zero
eigenvalue~\cite{Engelhardt:2002qs}. As vortices modify their percolation
behavior at the deconfinement transition, chiral symmetry is
restored~\cite{Engelhardt:2002qs}.

A recent study of double-winding Wilson loops~\cite{Greensite:2014gra} further favors the center vortex degrees of freedom as the dominating fluctuations in the QCD vacuum. The
spatial distribution of center vortex fields not only gives the area law falloff
for simple Wilson loops, but also shows the correct difference of areas behavior
for double-winding Wilson loops, in contrast to other confining gluonic field
fluctuations. Further details, including an outlook on the relation of the
vortex picture to other models of the strong interaction vacuum, can be found in
the reviews~\cite{Greensite:2003bk,Engelhardt:2004pf}. 

In the present work we revisit the random vortex world-surface model, which
will be introduced in the next section along with its main achievements, to
measure static meson-meson and baryon--anti-baryon correlators, as specified
in section~\ref{sec:meas} together with simulation details and the exponential
error reduction method. In these correlators, we observe bond rearrangement
behavior as the relative positions of the static quarks are varied, in
accordance with a strict minimal area law, operative already at finite
separations. The confining bonds within each mesonic or baryonic cluster
appear to be fully saturating, with no residual interactions between
clusters. Thus, cluster separation occurs not only asymptotically, where it
would be expected on general grounds, but already at intermediate distances.
We present these results in section~\ref{sec:res} and furthermore compare
them with theoretical expectations and measurements in lattice QCD. We draw
our conclusions in section~\ref{sec:con}.

\section{Random vortex world-surface model}\label{sec:model}

The random center vortex world-surface
model~\cite{Engelhardt:1999wr,Engelhardt:2003wm} describes the infrared
gluonic dynamics in the strong interaction vacuum in terms of collective
degrees of freedom called center vortices, which represent random closed
lines of quantized chromomagnetic flux. In terms of field configurations
in four-dimensional (Euclidean) space-time, these correspond to an ensemble
of closed random world-surfaces. The concrete implementation of this model
ensemble studied in the following employs a hypercubic lattice, on which
the random surfaces are composed of elementary lattice squares. The
ensemble is governed by an action related to surface
curvature\footnote{A systematic gradient expansion of the vortex
world-surface action starts with a Nambu-Goto term (proportional to
world-surface area), followed by a curvature term. The effects of these
terms are correlated; configurations with more area will generically
also contain more curvature. In~\cite{Engelhardt:1999wr,Engelhardt:2003wm},
the two-dimensional plane of coupling constants corresponding to the
aforementioned two action terms was explored, and it was indeed found
that the terms can to a certain extent be traded off against one
another. Lines of approximately constant physics were identified,
on which characteristics of $SU(2)$ and $SU(3)$ Yang-Mills theory are
reproduced. These lines of constant physics contain points on which
the Nambu-Goto term vanishes (but do not include points on which the
curvature term vanishes). Consequently, these points in the coupling
constant plane were adopted as the simplest realization of the
vortex world-surface model consistent with the confinement
characteristics of Yang-Mills theory.}: If two
of the aforementioned elementary squares composing a vortex surface share
a lattice link while lying in different lattice planes, an action increment
$c$ is incurred. This can be written formally in terms of a sum over
lattice links,
\begin{eqnarray}
S[q] &=& 
c \sum_x\sum_\mu \left[ \sum_{\nu < \lambda \atop \nu \neq \mu,
\lambda\neq \mu} \left( | q_{\mu\nu}(x) \, q_{\mu\lambda}(x) |
 + | q_{\mu\nu}(x) \, q_{\mu\lambda}(x-e_\lambda) | 
\right. \right. \label{curvature} \\
& & \ \ \ \ \ \ \ \ \ \ \ \ \ \ \ \ \ \
+ \left. | q_{\mu\nu}(x-e_\nu) \, q_{\mu\lambda}(x) |
 + | q_{\mu\nu}(x-e_\nu) \, q_{\mu\lambda}(x-e_\lambda) |
\right)\Bigg] \nonumber \\
&=& \frac{c}{2} \sum_{x}\sum_\mu \left[ \left[
\sum_{\nu\neq\mu} \left( | q_{\mu\nu}(x) | +
| q_{\mu\nu}(x-e_\nu) | \right) \right]^2 \! \! \! - \! \!
\sum_{\nu\neq\mu} \! \! \left[ | q_{\mu\nu}(x) | +
| q_{\mu\nu}(x-e_\nu) | \right]^2 \ \right] \ , \nonumber
\end{eqnarray}
where $q_{\mu \nu } (x)$ specifies the chromomagnetic flux associated with
the elementary square that extends from the site $x$ into the positive $\mu $
and $\nu $ directions. The variables $q_{\mu\nu}(x)$ take three values,
$q_{\mu \nu } (x) \in \{ -1,0,1 \} $; the value $0$ means that the
elementary square is not part of a vortex surface, whereas nonzero
quantized flux carried by the vortices is encoded in the other values
$\pm 1$. The sign of $q_{\mu\nu}(x)$ corresponds to the orientation of the
vortex flux, implying the relation $q_{\nu\mu}(x) = -q_{\mu\nu}(x)$,
as will be clear from the characterization of vortex flux by its effect
on Wilson loops circumscribing it, given in the next section. For the
purpose of evaluating Wilson loops, in the $SU(2)$ case it will be seen
that there is no physical distinction between the $\pm 1$ fluxes, and
one could equivalently adopt a scheme in which $SU(2)$ elementary vortex
squares only take the value $+1$; for consistency of notation, this
option is not employed here. In the $SU(3)$ case, by contrast, the $\pm 1$
fluxes have physically distinct effects on Wilson loops and one must
keep track of the orientation of vortex flux.

Vortex flux is subject to the Bianchi identity, i.e., it is continuous
modulo $2\pi $, {\it i.e.}, modulo Dirac strings~\cite{Engelhardt:2000wc,
Engelhardt:2002qs,Engelhardt:2003wm}. Vortex world-surfaces therefore
are closed, as already mentioned above. Nevertheless, in the $SU(3)$
case, vortex branchings are allowed; flux continuity is respected when
a $\pm 1$ vortex splits into two $\mp 1$ vortices, and vice versa. In the
$SU(2)$ case, such branchings do not occur; there is really only one
physical vortex flux, and the distinction between $+1$ and $-1$ flux is
purely formal for present purposes, as already noted above. This
corresponds to there being only one nontrivial center element in the
$SU(2)$ group. These constraints stemming from continuity of
flux must be respected in the generation of the random surface ensemble;
in practice, this is achieved by performing Monte-Carlo updates on all
six squares making up the surface of an arbitrary elementary
three-dimensional cube in the lattice simultaneously~\cite{Engelhardt:2003wm}.
Essentially, the continuous flux of a vortex of the shape of the
elementary cube surface is superposed onto the previously present flux.

This infrared effective model of strong interaction dynamics has been
explored extensively, ascertaining the reach of such a description of the
QCD vacuum in terms of vortex degrees of freedom with their simplified
effective action. Initial studies focused on the $SU(2)$ gauge group;
the random vortex world-surface model was seen to allow for both a
low-temperature confining as well as a high-temperature deconfined
phase~\cite{Engelhardt:1999wr}, which are separated by a second-order
phase transition~\cite{Engelhardt:2003wm}. The spatial string tension
in the deconfined phase~\cite{Engelhardt:1999wr}, the topological
susceptibility~\cite{Engelhardt:2000wc} and the (quenched) chiral
condensate~\cite{Engelhardt:2002qs} are predicted in quantitative
agreement with $SU(2)$ lattice Yang-Mills theory. Extending the model
to $SU(3)$ color, a weakly first-order deconfinement phase
transition~\cite{Engelhardt:2003wm} was found, and confirmed by
investigating the vortex free energy~\cite{Quandt:2004gy}. The baryonic
static potential exhibits a $Y$ law~\cite{Engelhardt:2004qq}, and also the
$SU(3)$ topological susceptibility was evaluated~\cite{Engelhardt:2010ft}.
Further gauge groups were studied in order to better understand the
systematics of competing confinement mechanisms~\cite{Holland:2003kg},
including $SU(4)$ color~\cite{Engelhardt:2005qu} and $Sp(2)$
color~\cite{Engelhardt:2006ep}. It was seen that the vortex picture
is sufficiently flexible to account for the confinement properties
associated with various gauge symmetries, provided the effective
dynamics are suitably tailored; in general, one must employ more
complex effective actions than the simple curvature action
(\ref{curvature}) describing the $SU(2)$ and $SU(3)$ cases.
The present work revisits the latter, widening the scope from the
simplest confinement properties studied previously, as described above,
to more complex static meson-meson and baryon--anti-baryon correlators.

\section{Observables and exponential noise reduction}\label{sec:meas}

Our objective is to evaluate expectation values of multiple Polyakov loop
operators in a center vortex background. Studies of quantities of this type
have also been suggested in~\cite{Cornwall:2009as}. Generally, Wilson
loops\footnote{A correlator of two Polyakov loops can be viewed as a
special case of a Wilson loop, extending along the entire temporal
direction of the lattice; note that no path ordering issues arise in the
present case, owing to the Abelian character of the $Z(N)$ vortex
configurations.} are influenced by the quantized chromomagnetic flux
carried by center vortices in a characteristic fashion:
For an arbitrary area spanning a given Wilson loop
(the choice of area being immaterial owing to flux continuity), each time
the area is pierced by a vortex world-surface\footnote{Wilson loops are
defined on the lattice which is dual to the one on which the vortex
world-surfaces are defined, i.e., on a lattice shifted by the vector
$(a/2, a/2, a/2, a/2)$, where $a$ is the lattice spacing. Consequently,
the notion of a Wilson loop area being pierced by a vortex world-surface
is unambiguous.}, the Wilson loop is multiplied by a phase corresponding
to a nontrivial center element of the gauge group. Specifically, in the
$SU(2)$ color case, each piercing contributes a phase $(-1)=\exp(\pm i\pi)$,
whereas in the $SU(3)$ color case, two distinct phase factors are possible,
corresponding to the two possible quantized vortex fluxes, namely,
$\exp(\pm i2\pi/3)$. Note, thus, that there is no physical distinction
in the $SU(2)$ case between the two orientations of vortex flux, as
already mentioned further above. To cast these properties into precise
language, in terms of the variables $q_{\mu\nu}(x)$, consider evaluating
an elementary Wilson loop, i.e., the plaquette $U_{\kappa\lambda}(y)$,
starting at the (dual) lattice site $y$, integrating first into the
positive $\kappa$ direction, then into the positive $\lambda$ direction,
and continuing around the plaquette. This plaquette is pierced (only) by
the lattice elementary square $q_{\mu\nu}(x)$, with the indices
$\kappa,\lambda,\mu,\nu$ spanning all four space-time dimensions and
$x=y+(\vec{e_\kappa}+\vec{e_\lambda}-\vec{e_\mu}-\vec{e_\nu})a/2$,
where $a$ denotes the lattice spacing. Thus, $U_{\kappa\lambda}(y)$ is
determined exclusively by $q_{\mu\nu}(x)$ as
\begin{equation}
U_{\kappa\lambda}(y)=\exp(i\pi/N\cdot\epsilon_{\kappa\lambda\mu\nu}q_{\mu\nu}(x))
\label{eqn:Wplaq}
\end{equation}
where $N$ is the number of colors and the usual Euclidean summation
convention over Greek indices applies. An arbitrary Wilson loop can be
evaluated~\cite{Engelhardt:2004qq} by finding a tiling of that loop by
a set of plaquettes, and multiplying the values (\ref{eqn:Wplaq}) of
those plaquettes in the given vortex configuration.

\begin{figure}[h]
	\centering
	a)\includegraphics[width=.4\linewidth]{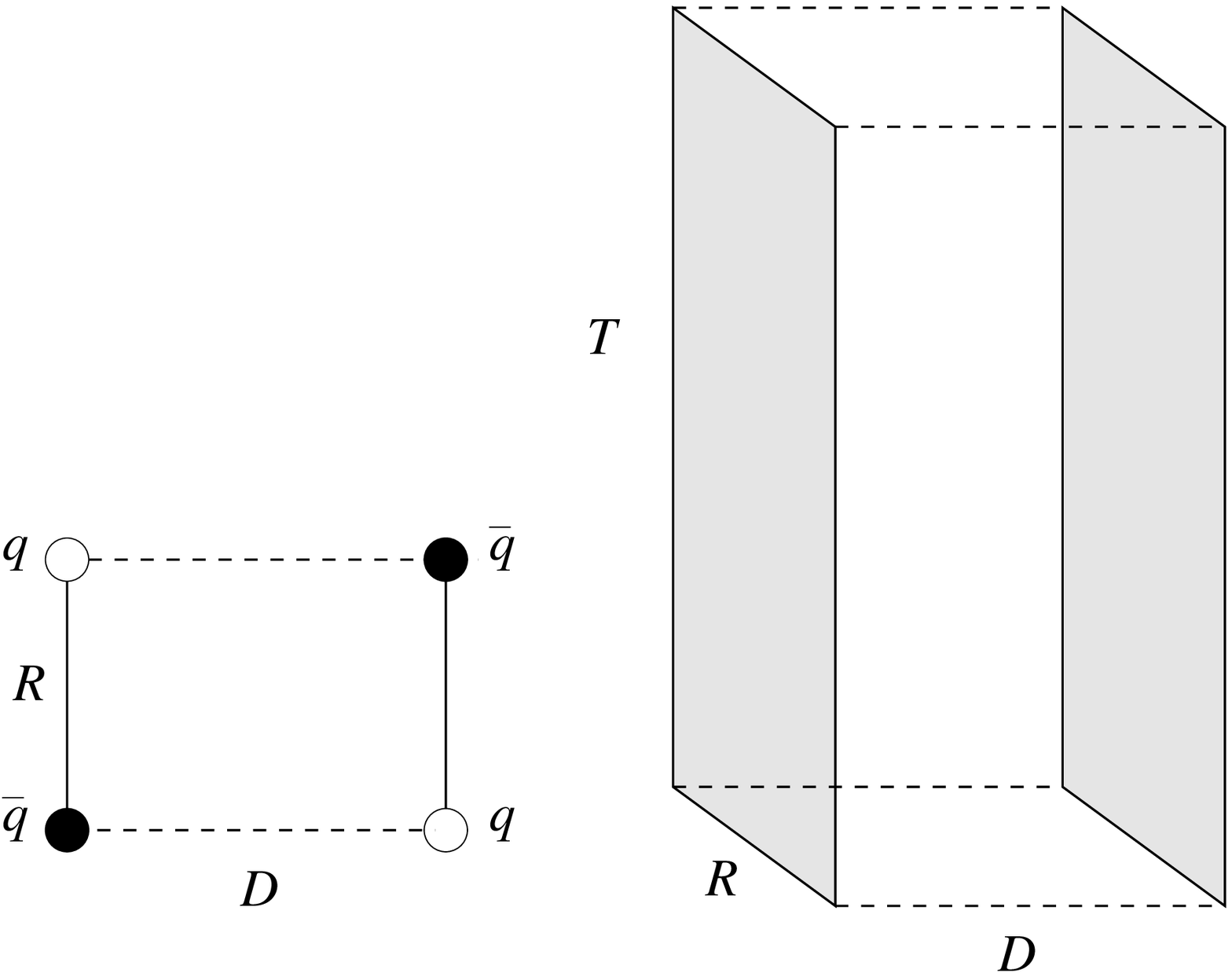}$\qquad$$\qquad$
	b)\includegraphics[width=.4\linewidth]{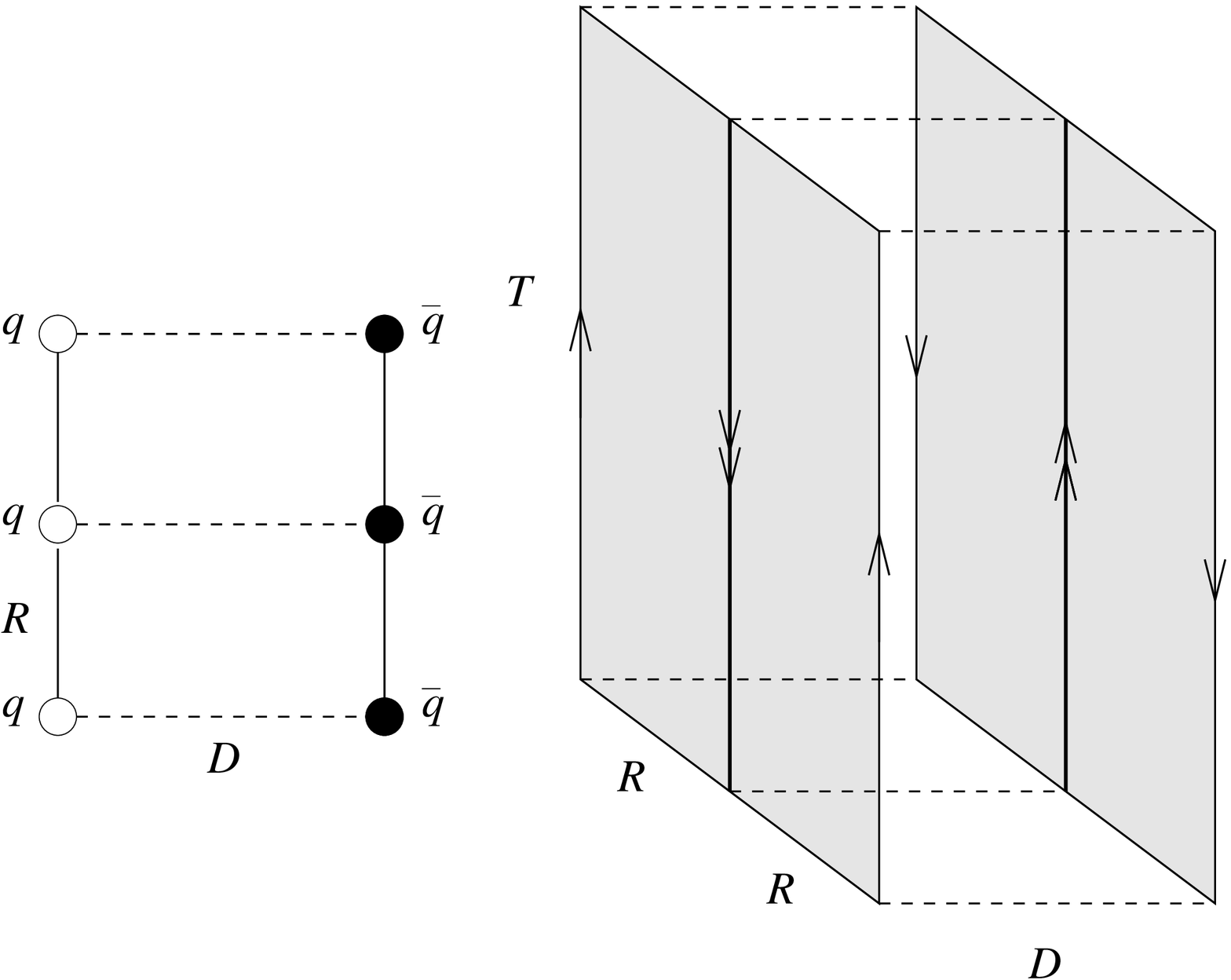}
	\caption{Polyakov loop setups for static a) meson and b) baryon correlators.}
	\label{fig:lps}
\end{figure}

To probe static meson-meson potentials, one measures the correlator of two
(flat) Wilson loops of size $R\times T$, {\it i.e.}, the quark and antiquark
in each static meson are separated by a distance $R$ in a spatial direction and
are propagating in time direction, with distance $D$ between the two mesons
in an orthogonal spatial direction. To probe static baryon--anti-baryon
potentials, the spatial setup depicted in Fig.~\ref{fig:lps}b was
adopted: Two Wilson loops of size $R\times T$ (static mesons) with opposite
orientation are combined to construct a propagating static linear baryon,
{\it i.e.}, the three quarks in the baryon lie equidistantly along one line
with total extent $2R$. This pair of Wilson loops is then again correlated
with a mirrored pair of loops for the anti-baryon at distance $D$. A
comment is in order concerning this particular choice for the spatial
arrangement of the quarks, which is owed to computational simplicity. A
realistic baryon of course is not described by any particular static
arrangement of quarks, but by a wave function which features a finite
probability density for linear, triangular, and a continuum of other
spatial arrangements of the quarks (presumably, approximately triangular
arrangements are somewhat more likely than approximately linear ones).
Studying all these possibilities in a representative manner lies beyond the
scope of this work. No particular choice will adequately represent an
actual baryon; this is not the intent of the present study. Instead, the
focus lies on the gluonic dynamics alone, namely, on the confining bond
rearrangements given a collection of quarks. To address this question, a
particular spatial arrangement is selected from among the continuum which
occurs in a baryon. The linear arrangement has the advantage that all bonds
occur along lattice axes, and therefore the resulting data do not have to be
disentangled with respect to, e.g., effects of rotational symmetry
breaking when bonds do not lie along axes. As already noted above,
a detailed study of the baryonic $Y$ law, including off-axis bonds,
was carried out in~\cite{Engelhardt:2004qq}.

In practice, in the present work, the Wilson loops were chosen to extend
over the entire temporal extent of the lattice, in which case the
correlators described above take the form of multiple Polyakov loop
correlators. Static meson-meson correlators are obtained from four,
static baryon--anti-baryon correlators from six Polyakov loops, see
Fig.~\ref{fig:lps}.

To reduce the numerical noise contaminating the mesonic and baryonic Polyakov
loop correlator measurements as far as possible, the noise reduction technique
introduced by L\"uscher and Weisz~\cite{Luscher:2001up} was employed, adapted
to the random center vortex world-surface model~\cite{Engelhardt:2004qq}:

As noted in conjunction with (\ref{curvature}), the vortex world-surface
action takes the form of a sum over contributions associated with lattice
links. The contribution at each link couples only elementary squares
attached to the link. The action can thus be decomposed in the form
\begin{equation}
S[q]\equiv S[q^t,q^s]=S^t[q^t]+\sum_i S^i[q^t,q_i^s]\ ,
\label{eqn:lwS}
\end{equation}
where the set of variables $q^t$ represents all elementary squares that
extend into the time and one space direction, whereas the set of variables
$q_i^s$ represents the elementary squares extending into two spatial
directions at fixed lattice time $i$. The $S^i$ piece of the action thus
contains the sum over spatial links at lattice time $i$, and $S^t$
contains the sum over all temporal links. The constraint of continuity
of flux can be treated in an analogous manner; it is a constraint that
must be satisfied at each lattice link separately~\cite{Engelhardt:2004qq},
and in each case involves precisely the elementary squares attached to
the link in question. The constraint therefore factorizes into terms
which either only couple the variables $q^t$ and $q^s_i$ at fixed lattice
time $i$, or couple only the variables $q^t$. One can thus express the
constraint of continuity of flux in the form
\begin{equation}
\delta[q]\equiv \delta[q^t,q^s]=\delta^t[q^t]\prod_i \delta^i[q^t,q_i^s]\ .
\label{eqn:lwd}
\end{equation}
Finally, Wilson loop observables obey a similar decomposition. As noted
above in conjunction with Eq.~(\ref{eqn:Wplaq}), a Wilson loop $W$ is
simply given as a product over the plaquettes making up a tiling of
the loop. Each of the plaquettes is determined exclusively by its dual
vortex elementary square via Eq.~(\ref{eqn:Wplaq}); one can therefore
group the product into factors as
\begin{equation}
W[q]\equiv W[q^t,q^s]=\prod_i W^i[q^t,q_i^s]\ .
\label{eqn:lwW}
\end{equation}
The noise reduction method given by L\"uscher and Weisz operates by
introducing a hierarchy into the Monte Carlo averaging of $W$ over
the variables $q$, governed by the action $S$ and the constraint $\delta$,
as follows,
\begin{equation*}
\langle W \rangle_{\{q;\delta;S\}}=\frac{1}{Z}\int[Dq^t]\delta^t[q^t]\exp(-S^t[q^t])\prod_i\int[Dq_i^s]\delta^i[q^t,q_i^s]W^i[q^t,q_i^s]\exp(-S^i[q^t,q_i^s])=\left\langle\prod_i\langle W^i\rangle_{\{q^s_i;\delta^i;S^i\}}
[q^t]\right\rangle_{\{q;\delta;S\}}
\end{equation*}
I.e., one keeps the variables $q^t$ fixed while performing the inner
averagings over the sets of variables $q_i^s$ according to the
corresponding actions $S^i$ and constraints $\delta^i$; then, the
outer expectation value over the product of these inner averages,
taken over the full set of variables $q$ according to the full action
$S$ and the full constraint $\delta$, may fluctuate much less than if
one were simply to average the original Wilson loop $W[q]$. As noted
in section~\ref{sec:model}, updates of the random vortex world-surfaces
are performed on all six surfaces of an elementary three-dimensional
cube in the lattice simultaneously; in the inner averaging step,
therefore, only cubes extending into three spatial directions, i.e.,
lying in a fixed time slice, are updated.

This algorithm can be straightforwardly iterated, i.e., further
levels of averaging can be introduced in which, at each level,
the quantities being averaged only depend on a subset of the variables
on which the quantities at the next-lower level depended. Concretely,
from the set of variables $q^t$ a further subset can be selected which
is kept fixed while one averages over its complement, provided the
action, the constraint and the observable can still be decomposed
analogously to Eqs.~(\ref{eqn:lwS})-(\ref{eqn:lwW}). In the present
work, one additional such iteration was performed, namely, the set
of variables $q^t$ was partitioned into the sets $q^t_{2i}$ of squares
extending from lattice time $(2i-1)$ to lattice time $(2i)$, with the
remaining squares constituting the set ̄$\bar q^t$ which is again kept fixed
at this second level in the hierarchy. This partitioning again yields
decompositions of the type~(\ref{eqn:lwS})-(\ref{eqn:lwW}), given that for
any $i$, the set of variables $q^t_{2i}$ only enters the product of
inner averages
\begin{equation*}
\langle W^{2i-1}\rangle_{{q^s_{2i-1};\delta^{2i-1};S^{2i-1}}}\langle
W^{2i}\rangle_{{q^s_{2i};\delta^{2i};S^{2i}}}\ .
\end{equation*}
In this second level of averaging, therefore, one performs updates
associated with all elementary three-dimensional lattice cubes except
for the ones extending from even lattice times $2i$ to the next higher
odd lattice times $2i+1$.

\iffalse
In the case of simple Wilson loops or Polyakov loop correlators, this method
practically gives an exponential noise reduction by exponentially enhancing the
number of measurements. 

\begin{figure}[h]
	\centering
	\includegraphics[width=.7\linewidth]{lw}
	\caption{Exponential noise reduction method for Wilson loops.}
	\label{fig:lw}
\end{figure}

For our meson-meson or baryon--anti-baryon correlators the
number of measurements is again exponentiated by a factor of two, 

\begin{figure}[h]
	\centering
	\includegraphics[width=.7\linewidth]{lw2}
	\caption{Exponential noise reduction method for Wilson loop correlators.}
	\label{fig:lw2}
\end{figure}
\fi

Finally, note that the auto-correlations between successive measurements
can be practically rendered negligible by updating the full configuration
a significant number of times before evaluating the next sample. This adds
only a small overhead to the total execution time, which is dominated by
the time required for the computation of the time-slice averages, but
exponentially reduced compared to measuring the observable with the same
accuracy but without the error reduction method~\cite{Luscher:2001up}.

\section{Measurements and Results}\label{sec:res}

We measure static meson-meson and baryon--anti-baryon correlators on $16^3\times4$ lattices,
which is still well in the confinement phase; the deconfinement transition occurs
between inverse temperature $T=2$ and $T=1$ for both $Z(2)$ and $Z(3)$ models
with optimized curvature action parameters, cf.~Eq.~(\ref{curvature}), namely,
$c=0.24$ and $c=0.21$ respectively~\cite{Engelhardt:1999wr,Engelhardt:2003wm}.
The corresponding string tensions are very similar, $\sigma a^2=0.755$ in the
$Z(2)$ case~\cite{Engelhardt:1999wr} and $\sigma a^2=0.766$ for the $Z(3)$
vortex model~\cite{Engelhardt:2004qq}, implying a lattice spacing of about
$a=
\iffalse
197.327\sqrt{\sigma a^2}/440=
\fi
0.39$ fm in both cases if one sets the zero-temperature string tension to
$\sigma=(440MeV)^2$. In practice, it turned out to be efficient 
to carry out the innermost averaging in the error reduction method by L\"uscher 
and Weisz~\cite{Luscher:2001up} described above using 8000 configurations; 
in the second-level averaging, 800 configurations were used. For the outermost
averaging, typically 200 configurations were enough to achieve a sufficient
level of accuracy. The correlator eventually becomes sensitive to the double-precision
machine accuracy at $-log\langle W(R,T=4)_xW(R,T=4)_{x+D}\rangle/T=6$,
{\it i.e.}, for individual meson/baryon loops $\langle W(R,T=4)\rangle\approx10^{-12}$. 

\subsection{Static meson-meson correlator in Z(2) and Z(3) vortex background}

\begin{figure}[h]
	\centering
	a)\includegraphics[width=.48\linewidth]{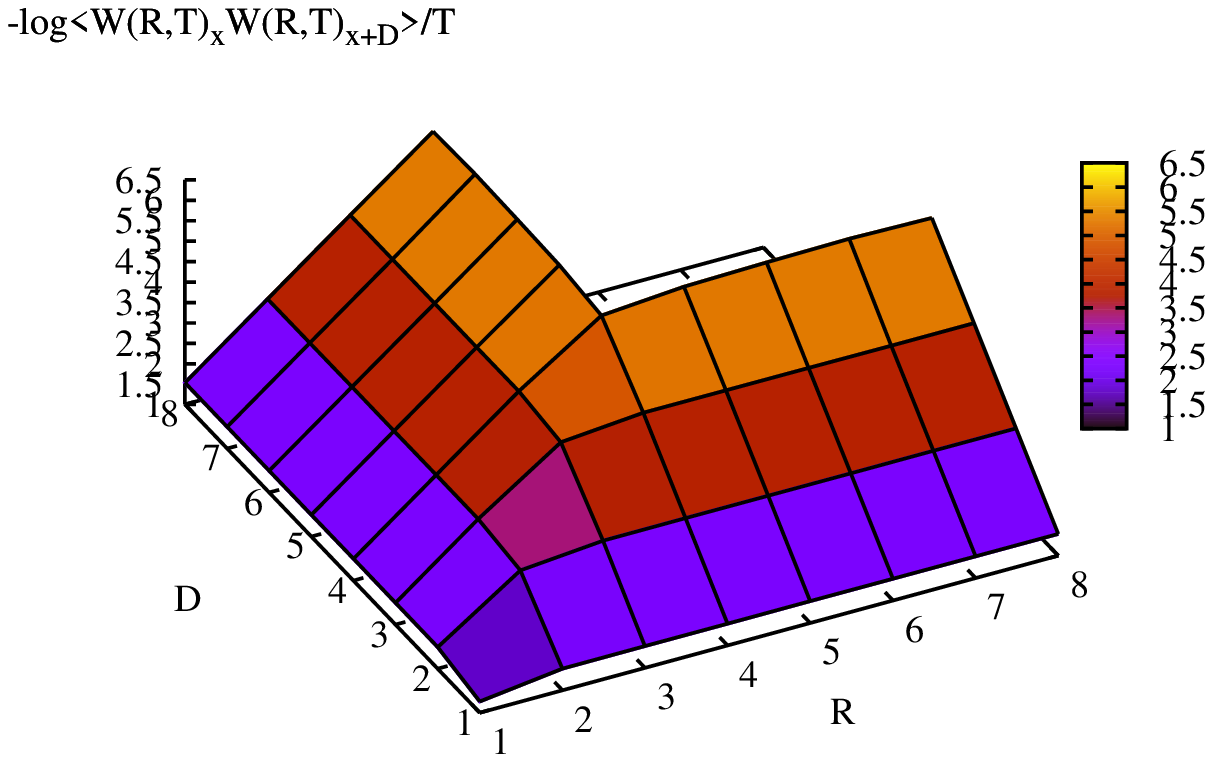}$\qquad$
	b)\includegraphics[width=.4\linewidth]{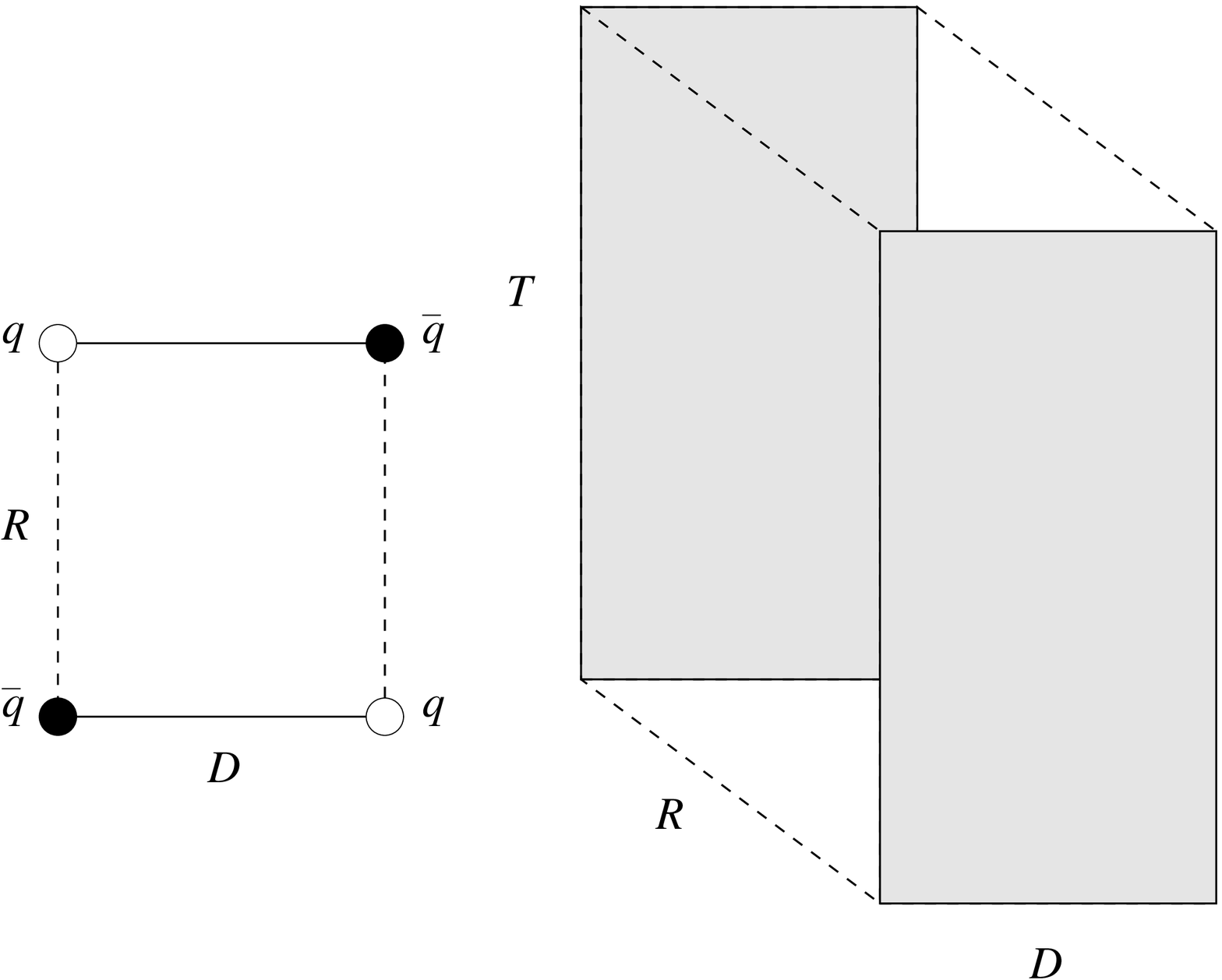}
	\caption{a) Z(2) static meson-meson Polyakov loop correlators $-log\langle
	W(R,T=4)_xW(R,T=4)_{x+D}\rangle/T$ for various distances $R$ between quark
	and anti-quark and distances $D$ between static mesons on
	$16^3\times4$ lattices (the Z(3) results look, apart from the slightly
	different string tension, the same). b) The expectation values of the
	static meson-meson correlator are given by the minimal area of possible
	Polyakov loop combinations and display bond rearrangement between the quarks if
	$D<R$, compare to Fig.~\ref{fig:lps}a.}\label{fig:m23}
\end{figure}

\begin{figure}[h]
	\centering
	a)\includegraphics[width=.44\linewidth]{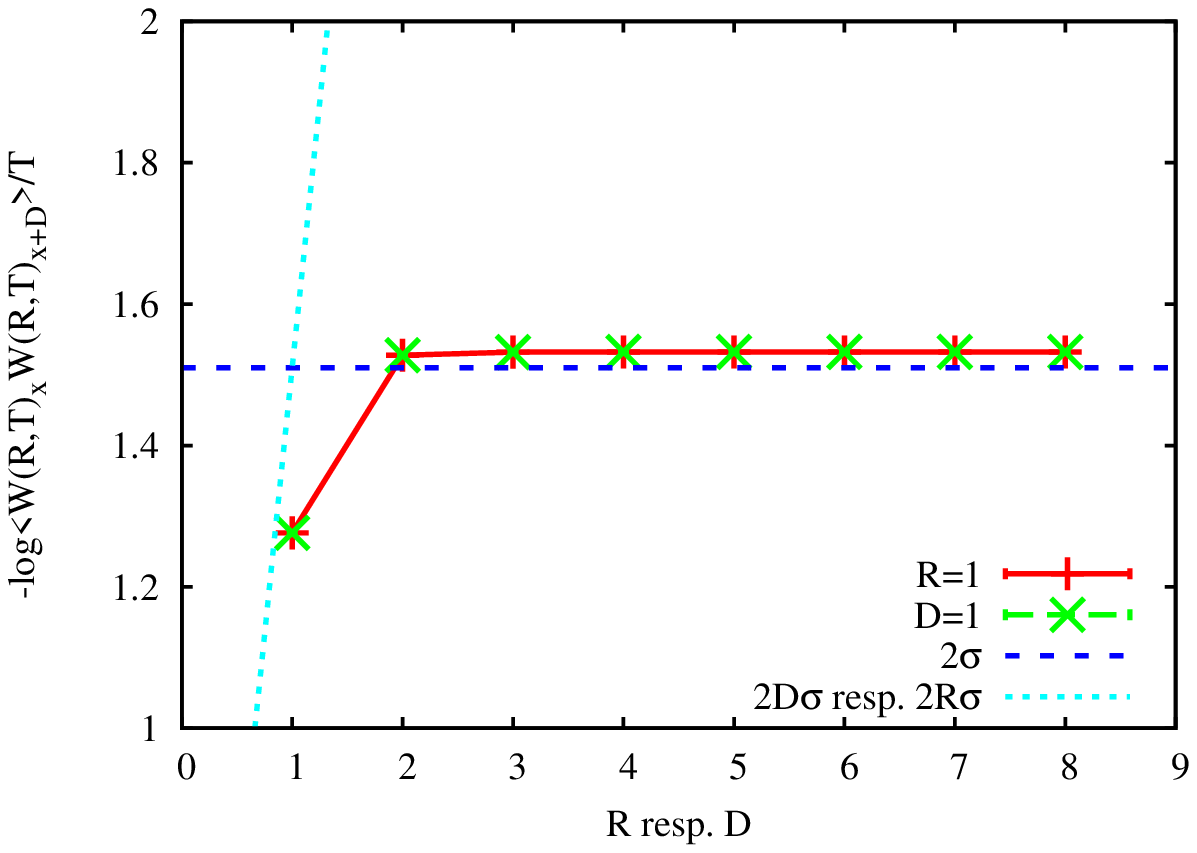}
	b)\includegraphics[width=.44\linewidth]{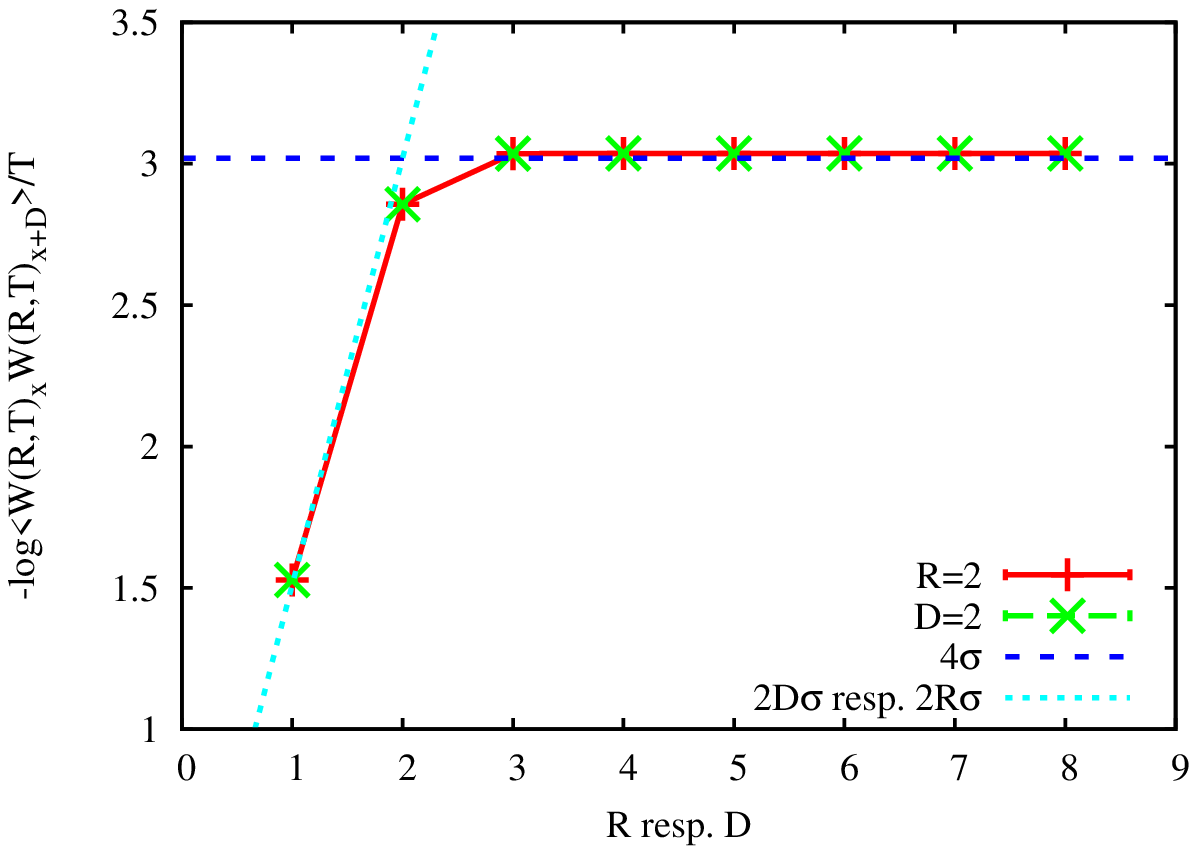}\\
	c)\includegraphics[width=.44\linewidth]{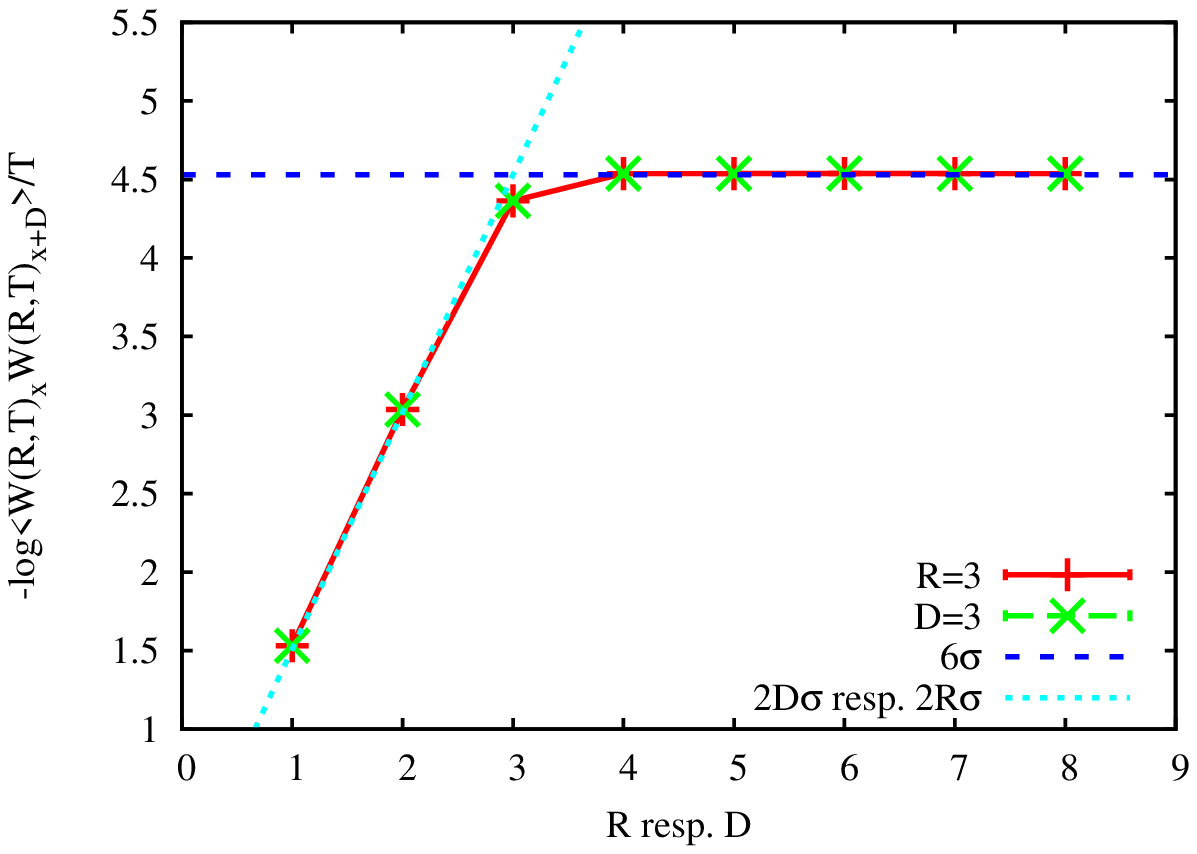}
	d)\includegraphics[width=.44\linewidth]{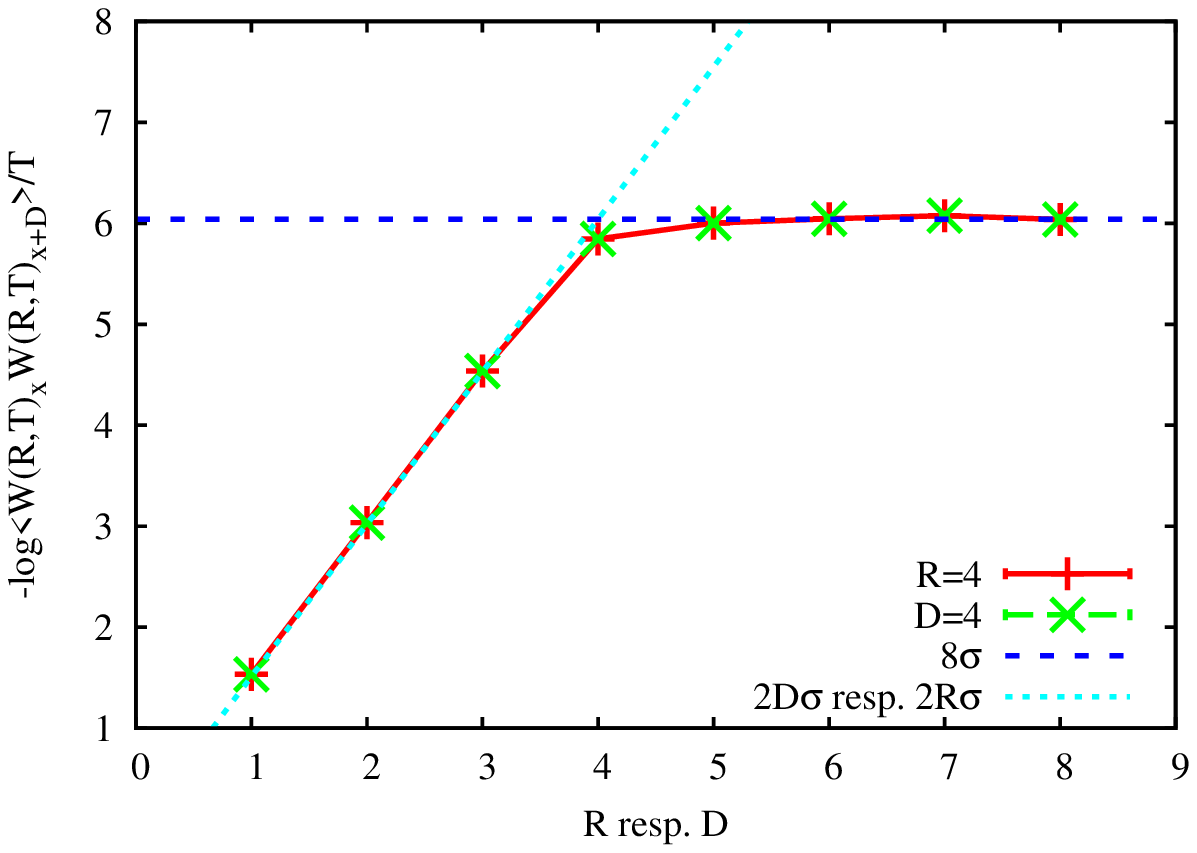}\\
	e)\includegraphics[width=.44\linewidth]{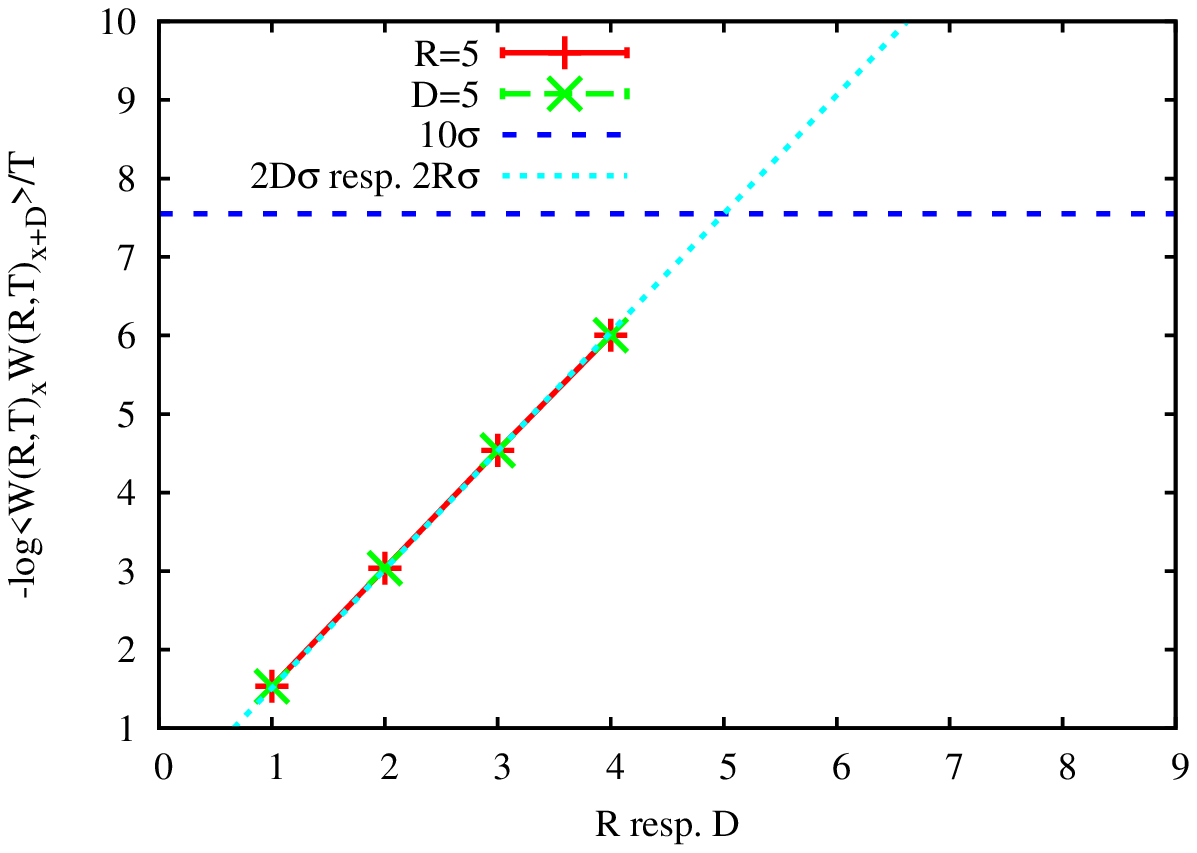}
	f)\includegraphics[width=.44\linewidth]{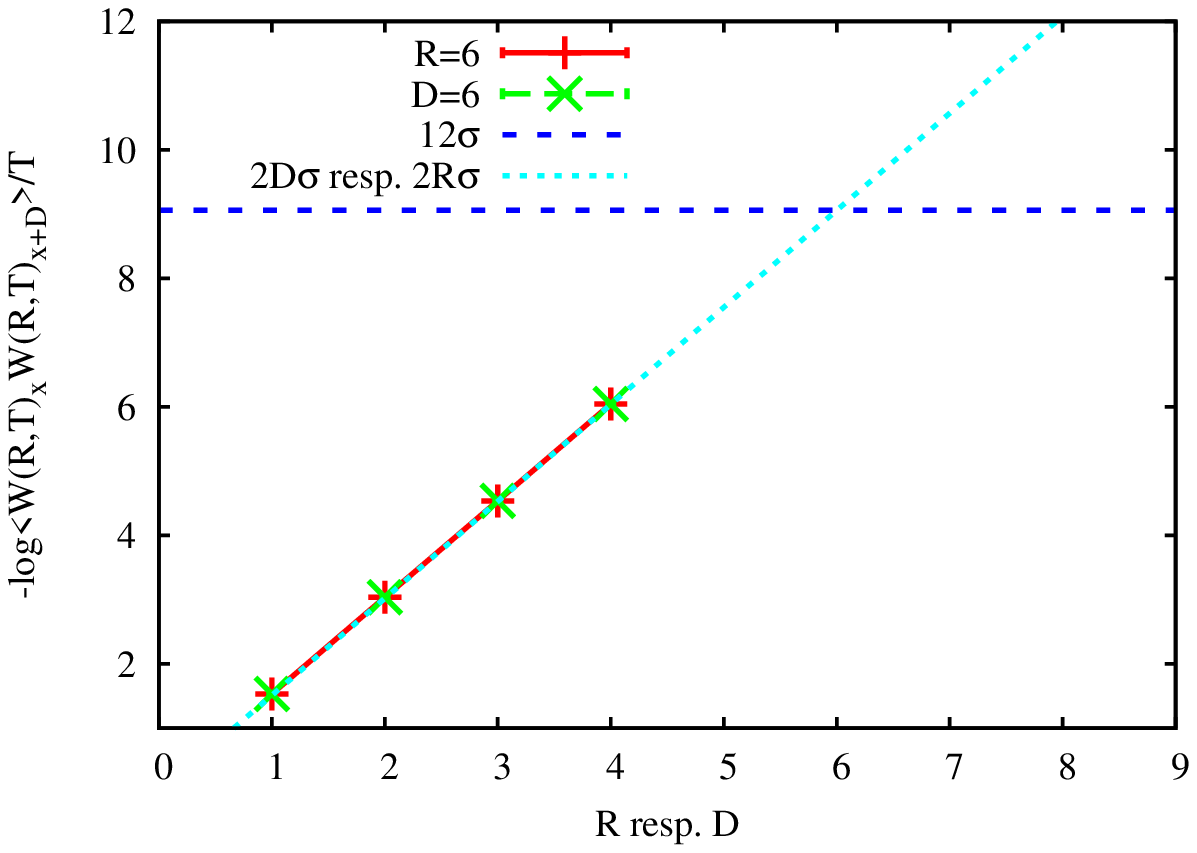}\\
	g)\includegraphics[width=.44\linewidth]{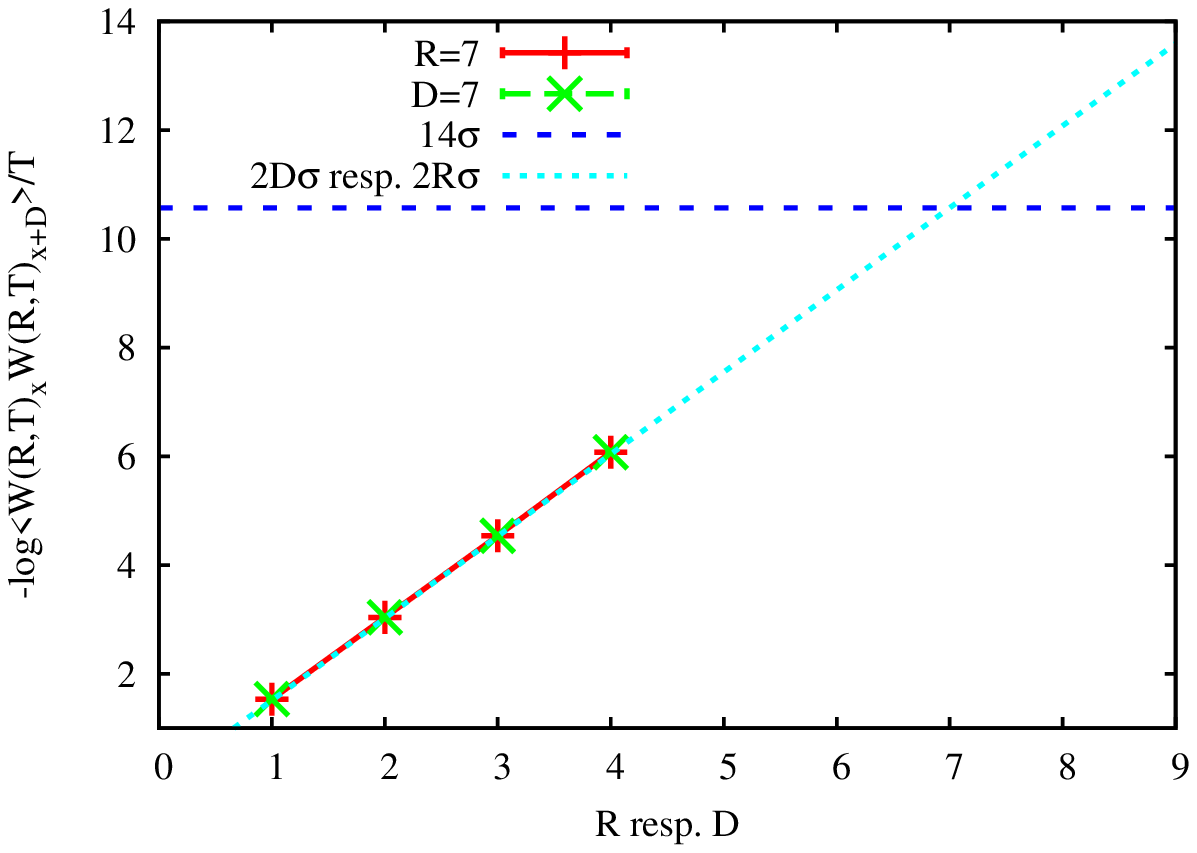}
	h)\includegraphics[width=.44\linewidth]{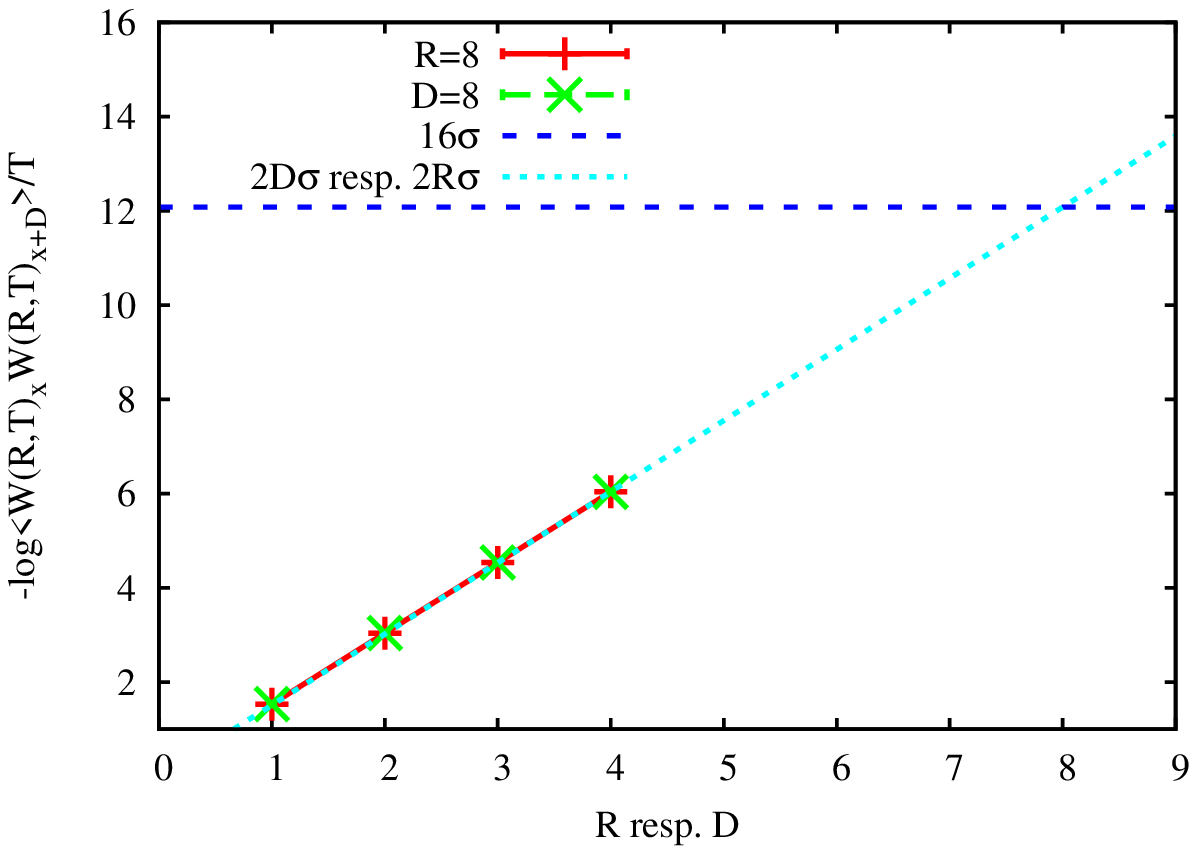}
	\caption{Z(2) static meson-meson potentials (Polyakov loop correlators)
$-log\langle W(R,T=4)_xW(R,T=4)_{x+D}\rangle/T$ for various distances $R$
between quark and anti-quark and distances $D$ between static mesons on
$16^3\times4$ lattices. Range of extracted lattice data is limited by
machine precision.}
	\label{fig:mz2}
\end{figure}

\begin{figure}[h]
	\centering
	a)\includegraphics[width=.44\linewidth]{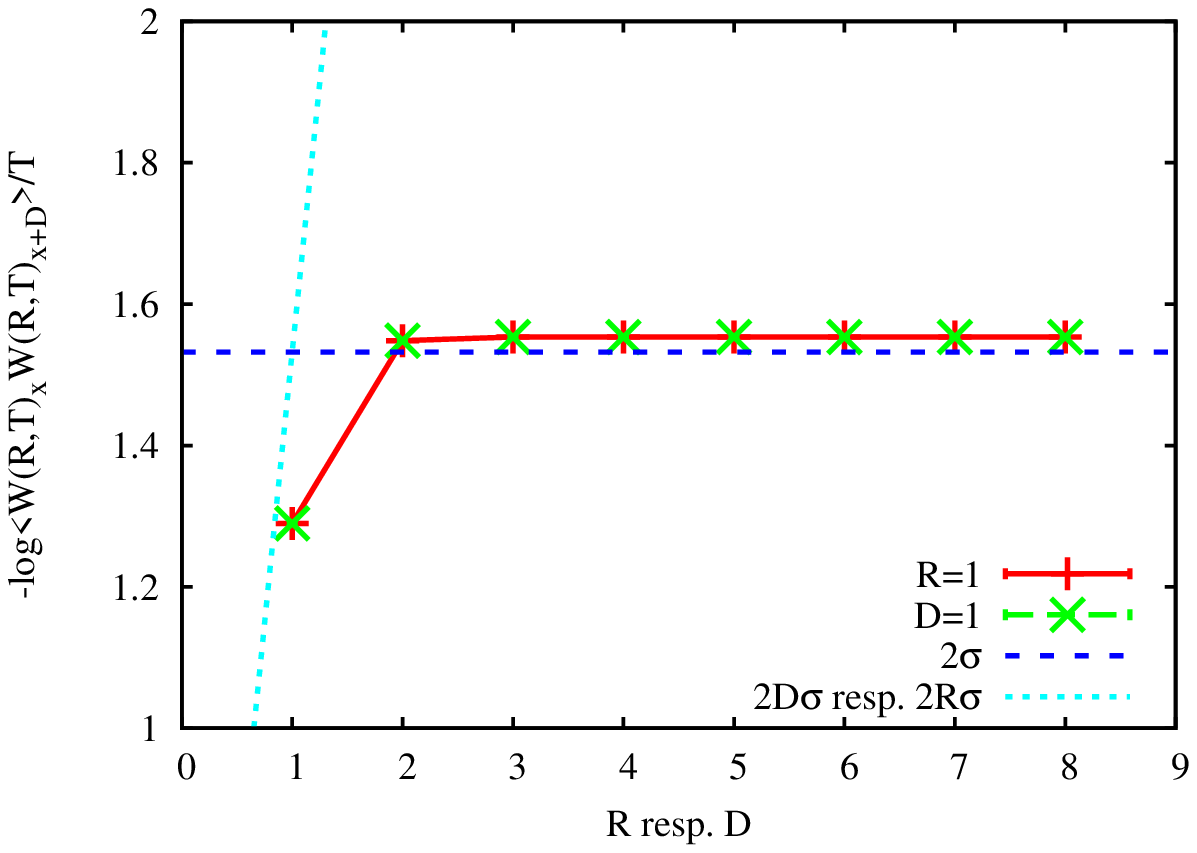}
	b)\includegraphics[width=.44\linewidth]{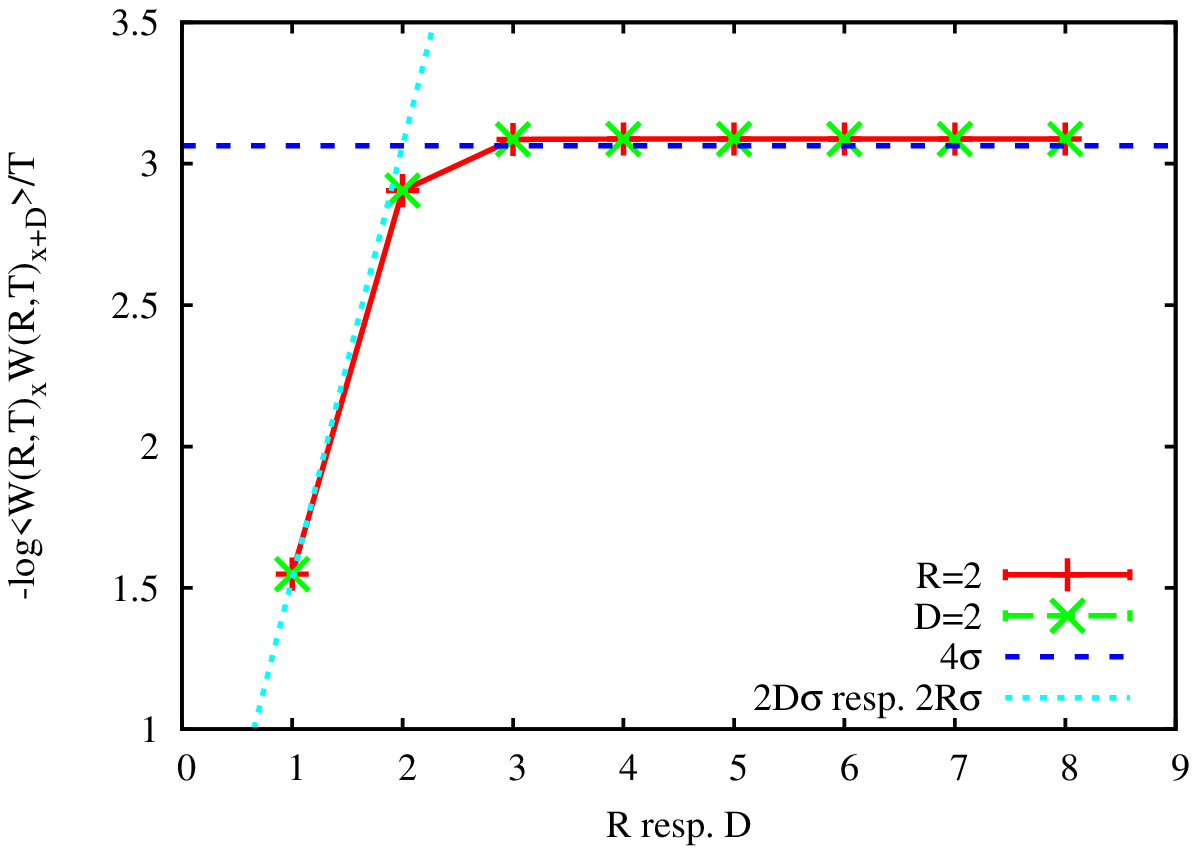}\\
	c)\includegraphics[width=.44\linewidth]{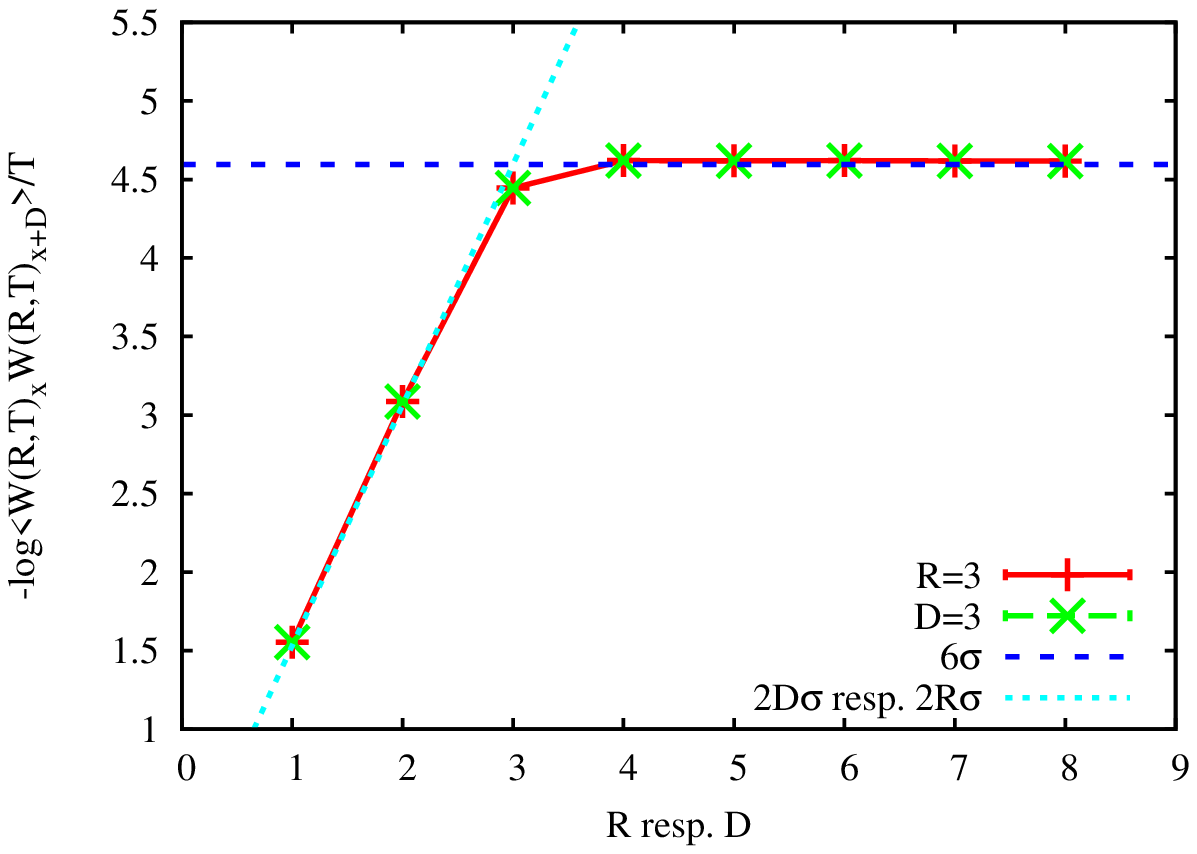}
	d)\includegraphics[width=.44\linewidth]{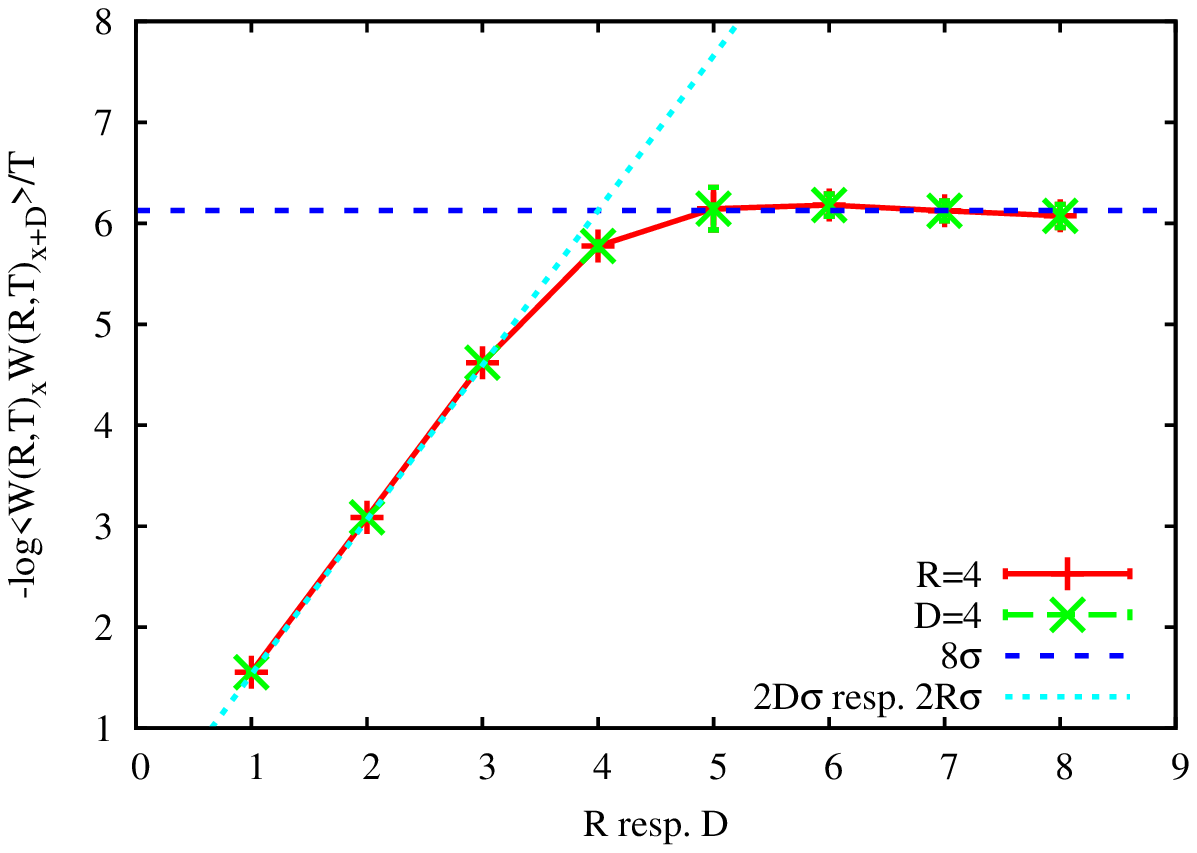}\\
	e)\includegraphics[width=.44\linewidth]{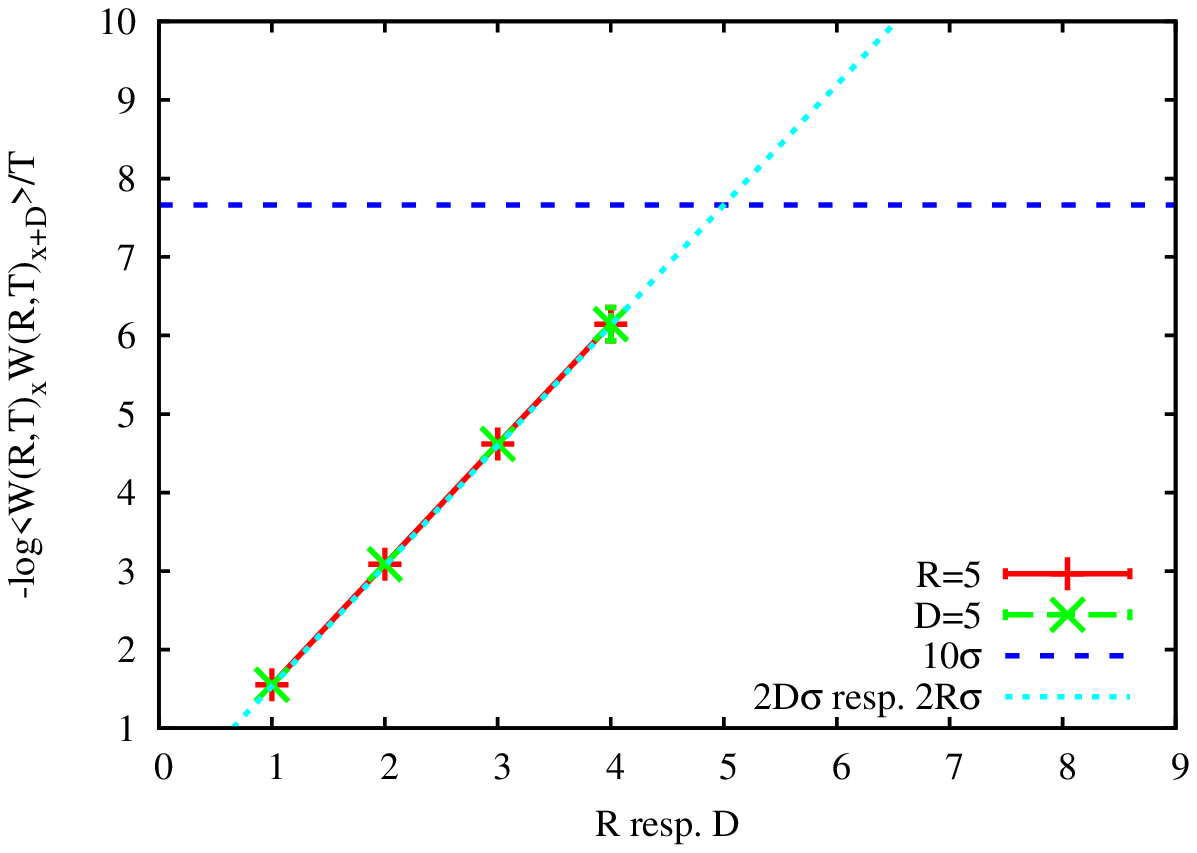}
	f)\includegraphics[width=.44\linewidth]{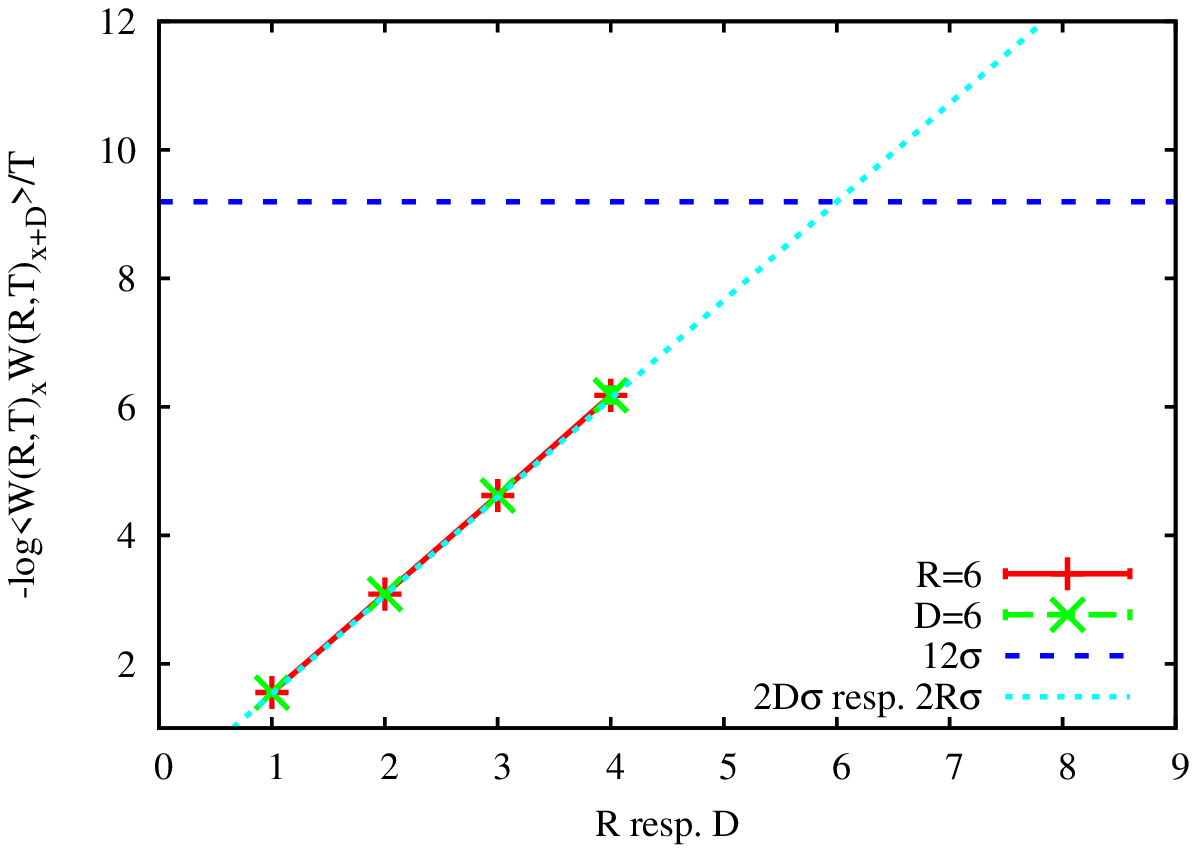}\\
	g)\includegraphics[width=.44\linewidth]{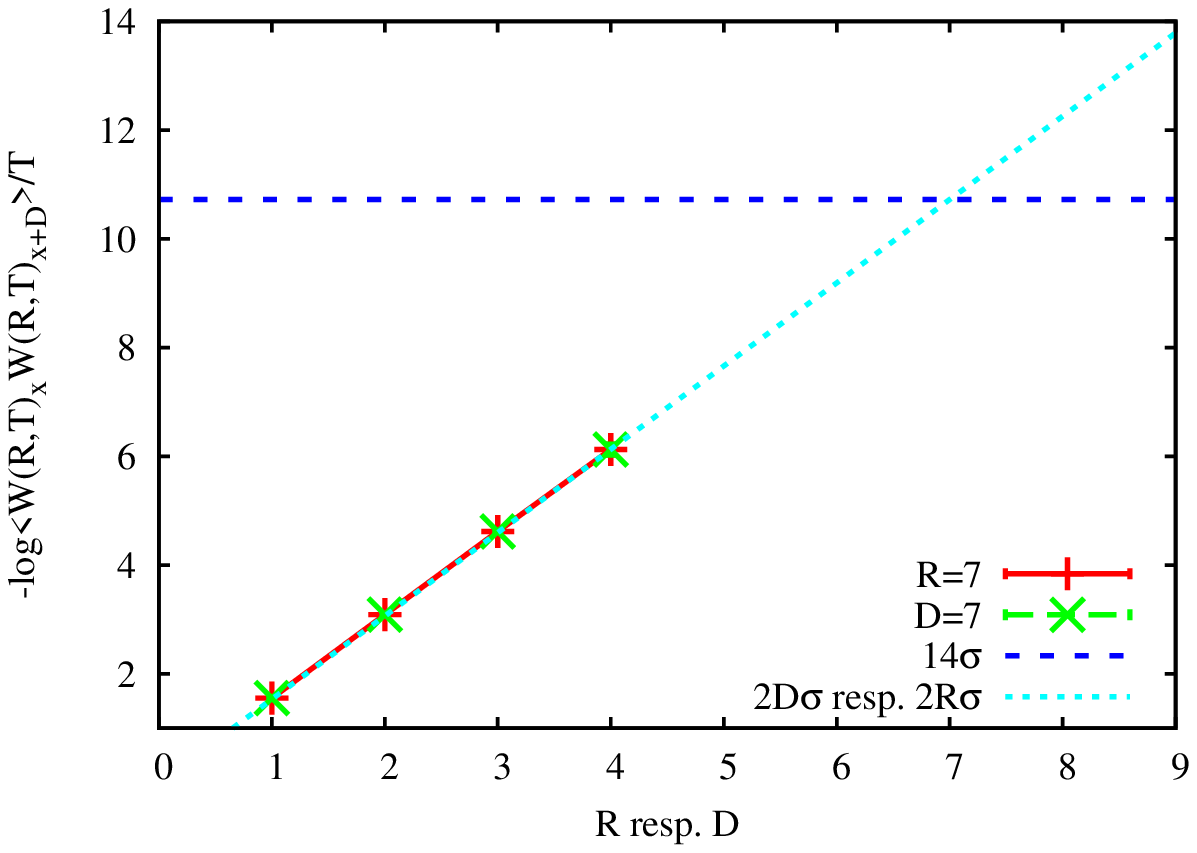}
	h)\includegraphics[width=.44\linewidth]{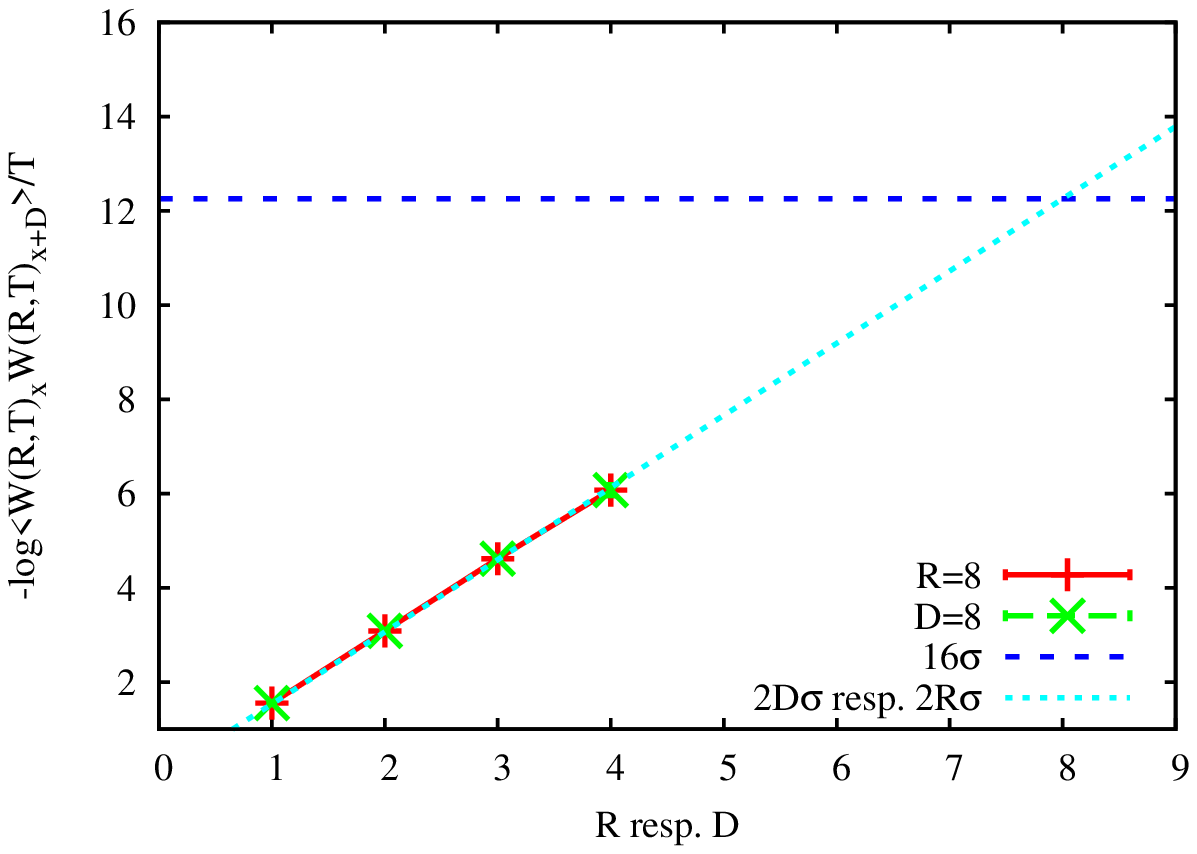}
	\caption{Z(3) static meson-meson potentials (Polyakov loop correlators)
$-log\langle W(R,T=4)_xW(R,T=4)_{x+D}\rangle/T$ for various distances $R$
between quark and anti-quark and distances $D$ between static mesons on
$16^3\times4$ lattices. Range of extracted lattice data is limited by
machine precision.}
	\label{fig:mz3}
\end{figure}

We show the static meson-meson potential
$-log\langle W(R,T=4)_xW(R,T=4)_{x+D}\rangle/T$ for various quark source
separations $R$ and distances $D$ between two static mesons in
Fig.~\ref{fig:mz2} for the $Z(2)$ case and in Fig.~\ref{fig:mz3} for the
$Z(3)$ vortex model. The potential is in fact symmetric with respect to
$R$ and $D$; the plots for constant $R$ and $D$, respectively, hence lie
on top of each other\footnote{It should be noted that this $R-D$ symmetry
is manifest in the multiple Polyakov loop correlators which are being
evaluated in the present specific calculational setup; on the other hand,
for general Wilson loop correlators which do not extend over the entire
temporal extent of the lattice, one would presumably have to take care
to allow for sufficiently long Euclidean time evolution in order to render
switching on-and-off effects, which break the symmetry, negligible, and
thus recover the $R-D$ symmetry.}. Once
the distance $D$ between the two static mesons becomes smaller than the quark
source separation within the mesons, {\it i.e.}, the spatial meson loop extent
$R$, we observe bond rearrangement, {\it i.e.}, the chromoelectric strings will
always connect quark sources such as to minimize their combined length. The
potential follows a direct area law as predicted by vortex models; the bond
rearrangement just means that the correlator will always measure the minimal
area of possible Polyakov loop combinations, and its values can be easily
reproduced by the form $2R\sigma$ or $2D\sigma$ respectively, indicated by blue
(dashed) and cyan (short dashed) lines using the corresponding string tensions
from above. As the two string tensions are almost the same, also $Z(2)$ and
$Z(3)$ results are likewise and we only show the
$Z(2)$ correlator as a 3D plot in Fig.~\ref{fig:m23}a vs. $R$ and $D$,
reflecting the $R-D$ symmetry, and the bond rearrangement, {\it i.e.},
minimal area behavior. As long as machine precision is sufficient,
{\it i.e.}, up to distances $R=4$ or $D=4$, the potentials nicely follow
the cyan (short dashed) and blue (dashed) predictions of minimal
area: the potential first scales linearly with quark (meson) separation and
stays constant once the active bonds are not extended anymore. At bond
rearrangement regions where $R\approx D$ the data show small deviations
which are most likely caused by mixed states. At small separations,
cf.~Figs.~\ref{fig:mz2}a,b and~\ref{fig:mz3}a,b, the potentials are still
affected by short-distance modifications to the asymptotic linear behavior,
and thus they lie slightly above the constant blue (dashed) line.

Defining the potential energy $V(R)$ to asymptotically go to zero, {\it i.e.}
$V(R)=E(R)-E(\infty)$, we have to shift our potentials by their asymptotic
values $2R\sigma$, {\it i.e.}, the two mesonic areas, in order to compare our
results with $SU(2)$ and $SU(3)$ lattice studies~\cite{Fiebig:2005sw,Bali:2010xa}.
Taking into account the fixed lattice spacing $a=0.39$ fm of the vortex models and
the fact that center vortices do not reproduce Coulomb effects, the potentials
are reproducing QCD results including the bond rearrangement
behavior~\cite{Lenz:1985jk}, compare for example Fig.~\ref{fig:mz3}b and Fig. 1b
in~\cite{Bali:2010xa}.

\subsection{Static baryon--anti-baryon correlator in Z(3) vortex background}

\begin{figure}[h]
	\centering
	a)\includegraphics[width=.44\linewidth]{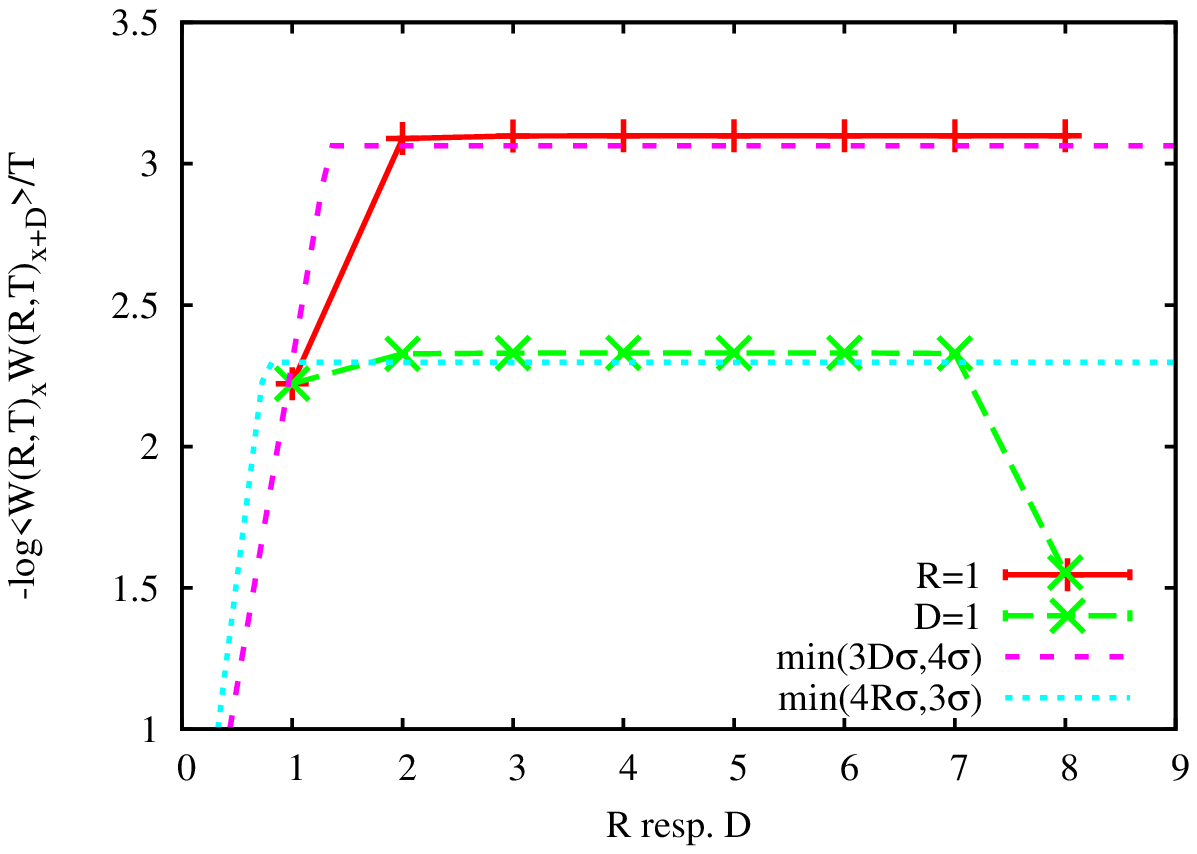}
	b)\includegraphics[width=.44\linewidth]{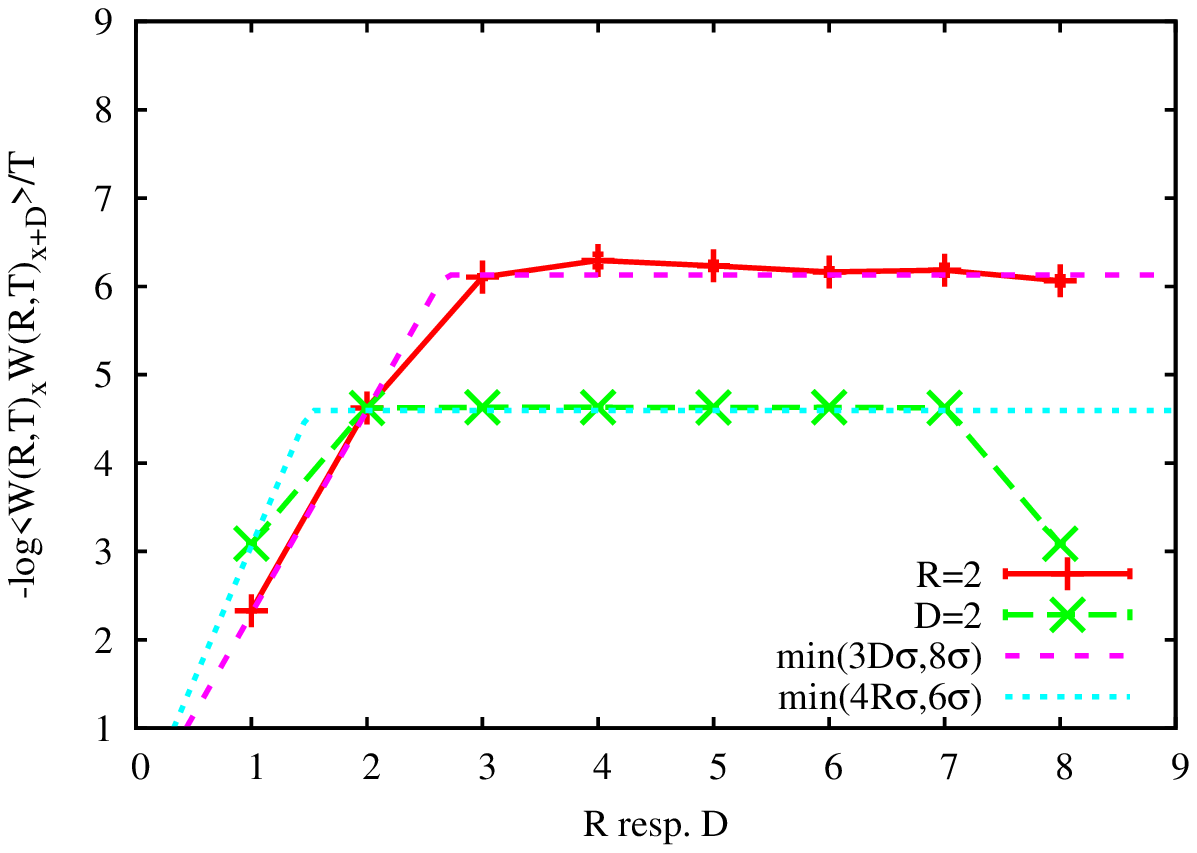}\\
	c)\includegraphics[width=.44\linewidth]{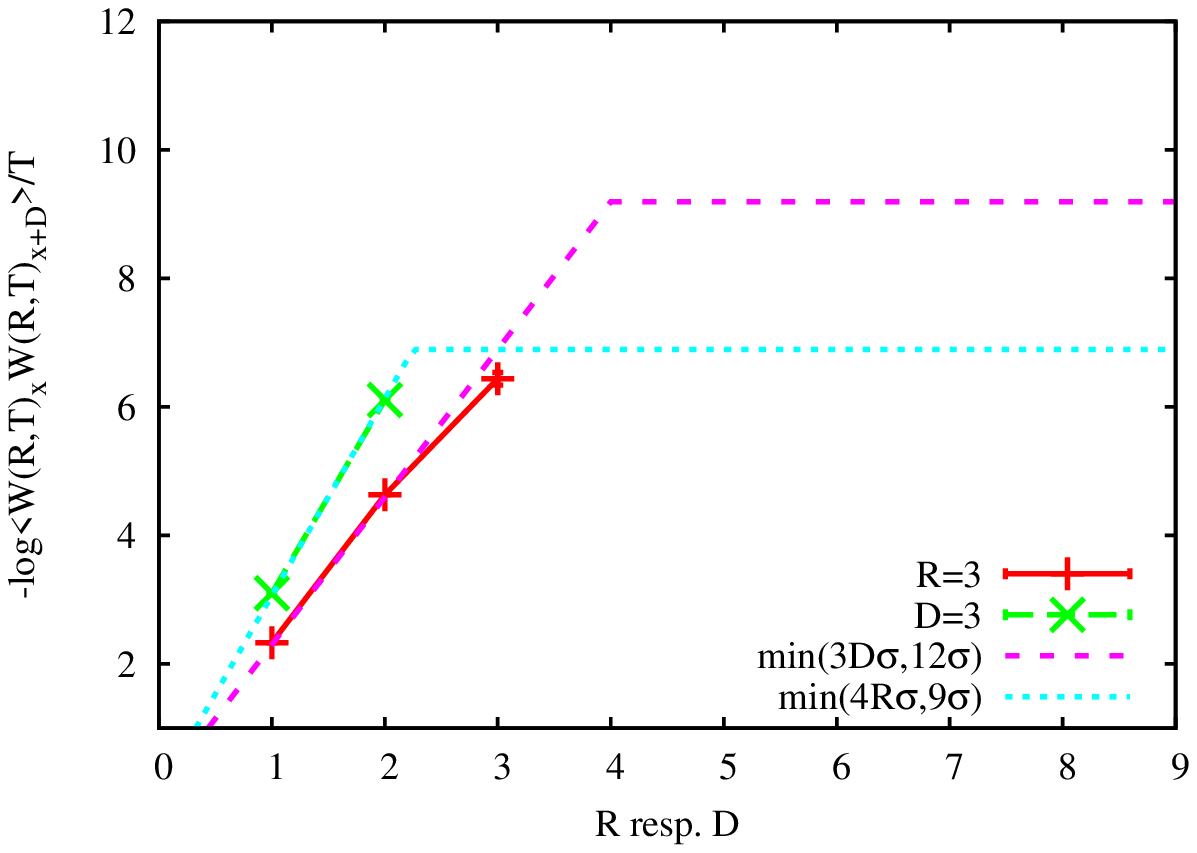}
	d)\includegraphics[width=.44\linewidth]{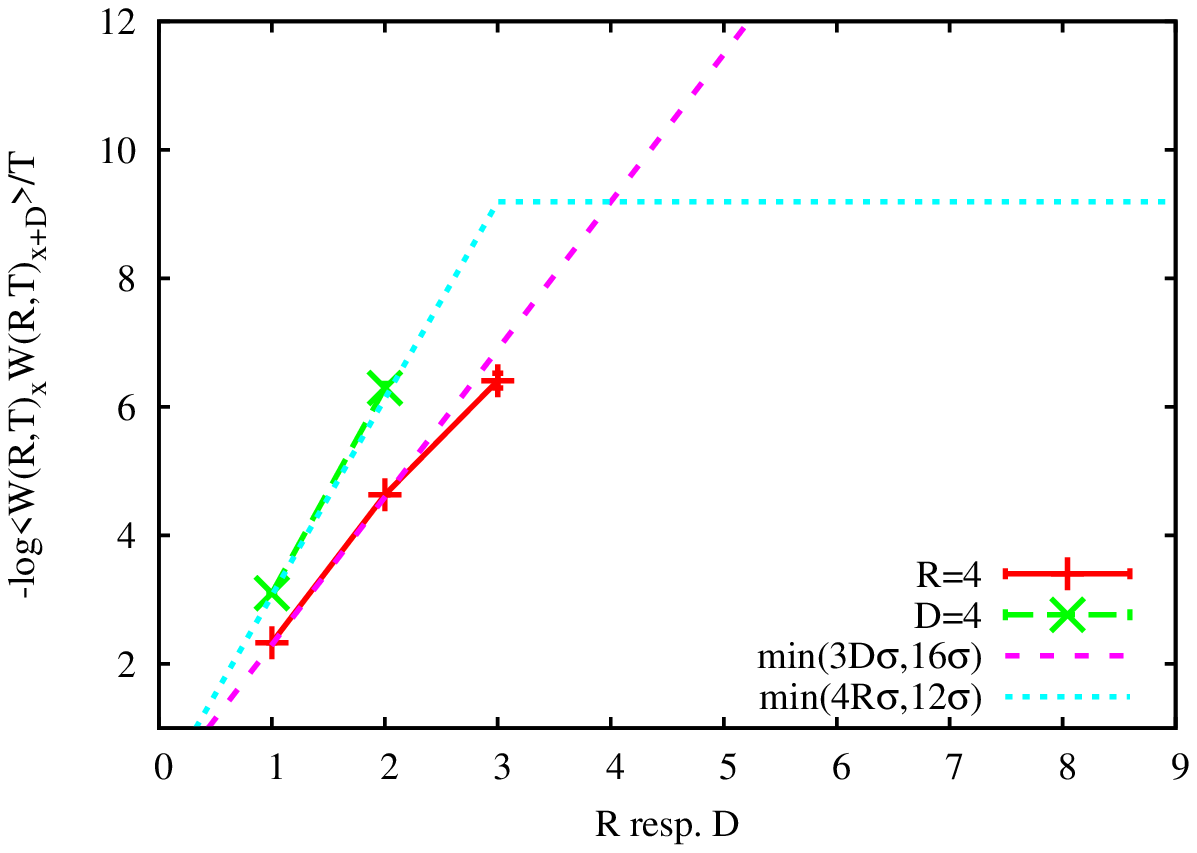}\\
	e)\includegraphics[width=.44\linewidth]{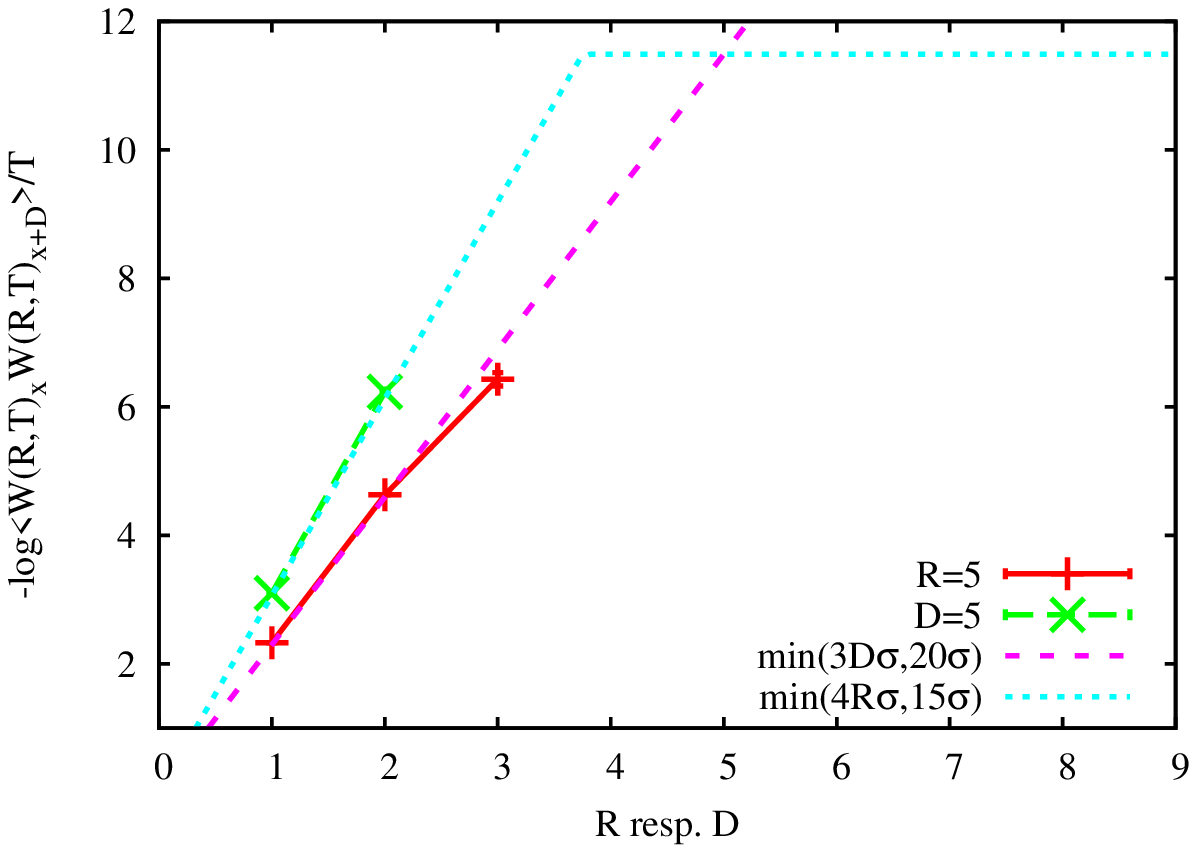}
	f)\includegraphics[width=.44\linewidth]{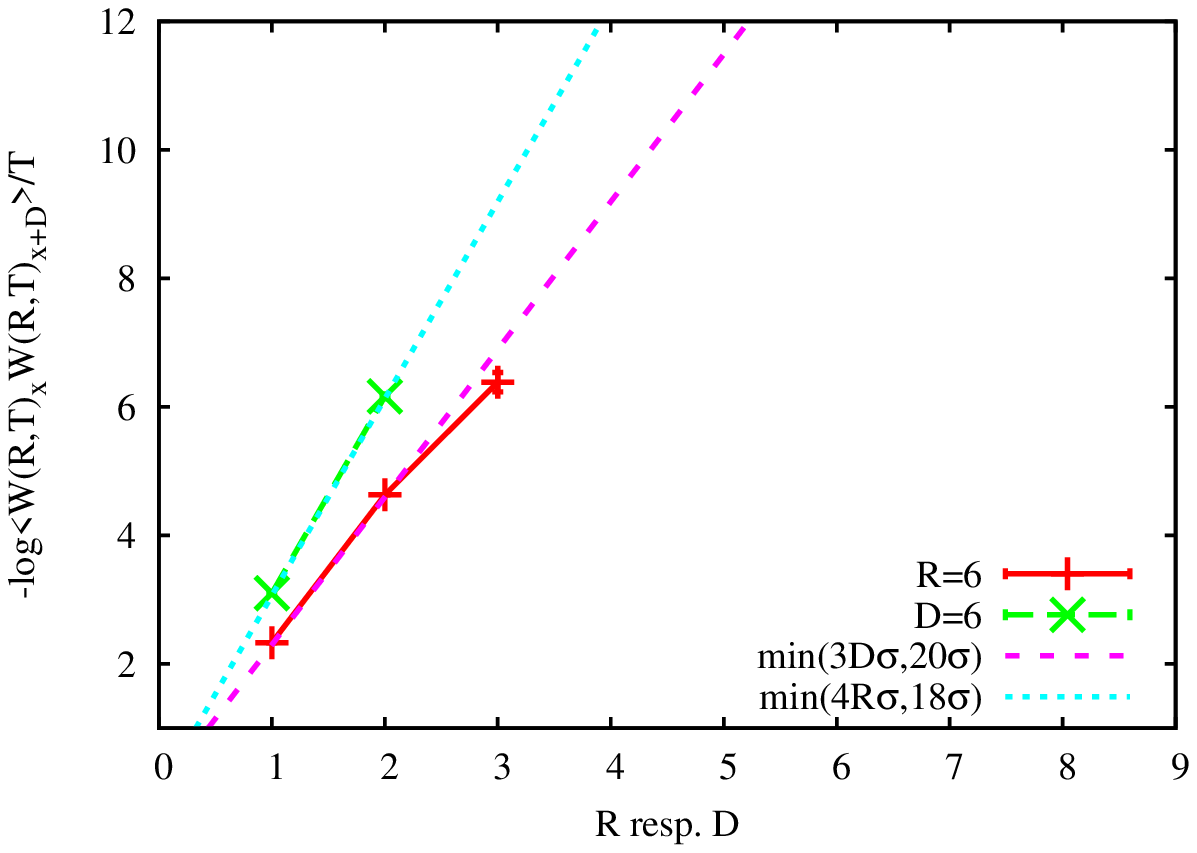}\\
	g)\includegraphics[width=.44\linewidth]{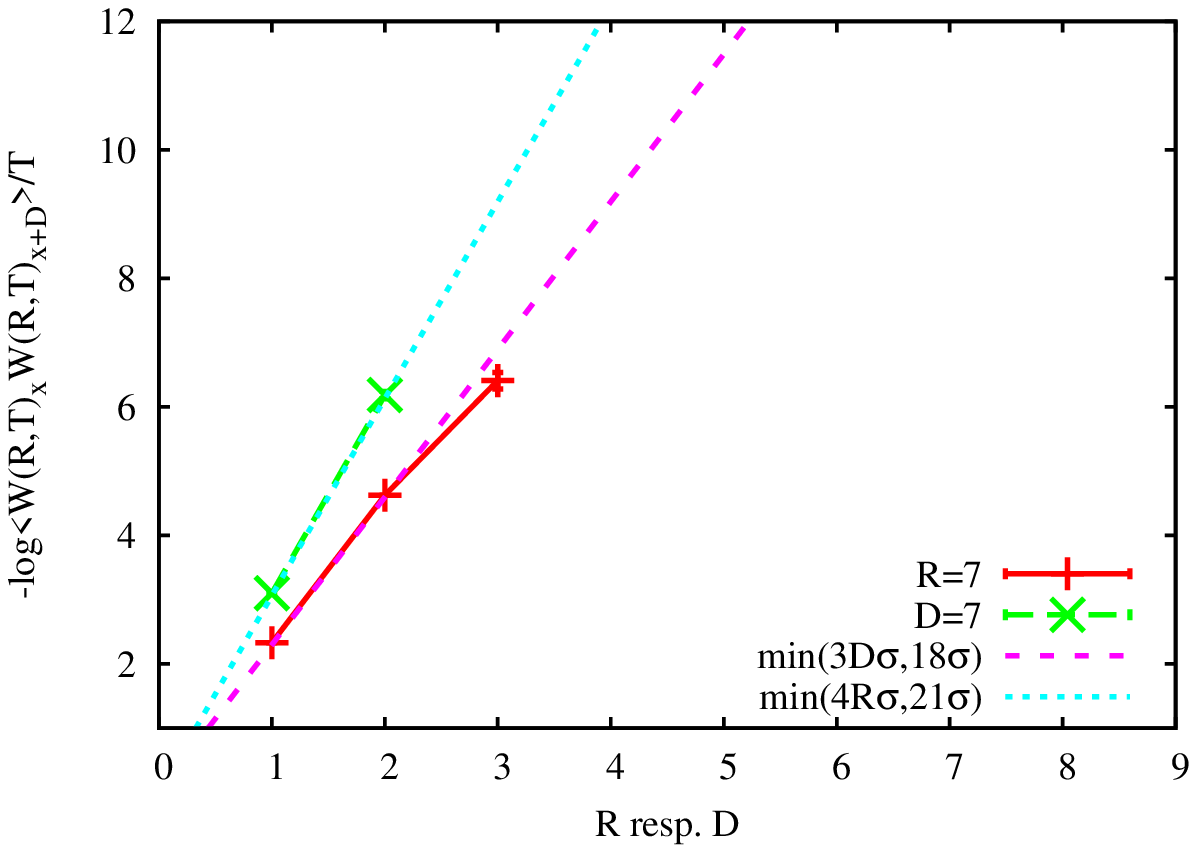}
	h)\includegraphics[width=.44\linewidth]{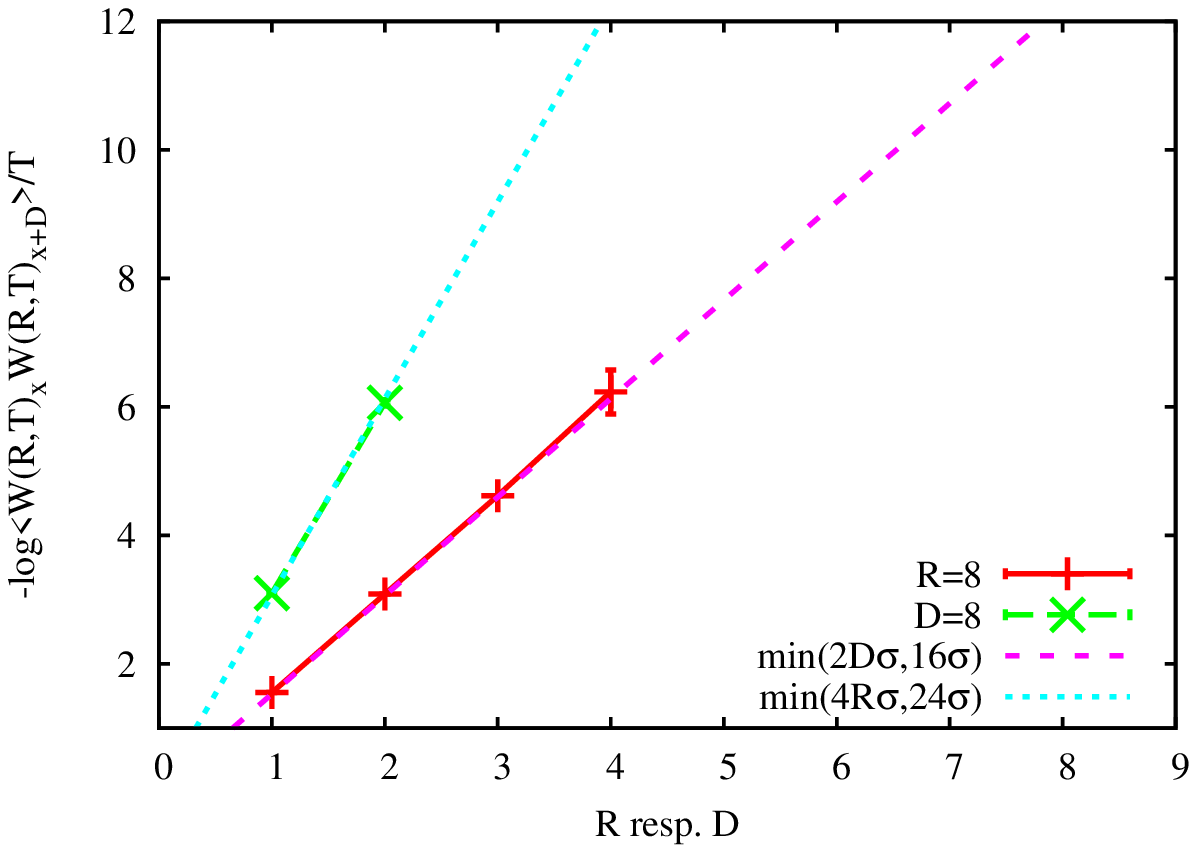}
\caption{$Z(3)$ static baryon--anti-baryon potentials (Polyakov loop
correlators) $-log\langle W(R,T=4)_xW(R,T=4)_{x+D}\rangle/T$ for various
distances $R$ between quark sources within the static baryons and distances $D$
between static baryon and anti-baryon on $16^3\times4$ lattices, compared with
minimal area law predictions in cyan \hfill (short dashed) \hfill for constant $D$ data \hfill
(green, dashed)}
\label{fig:wcb}
\end{figure}

\begin{figure}[h]
	\centering
	a)\includegraphics[width=.5\linewidth]{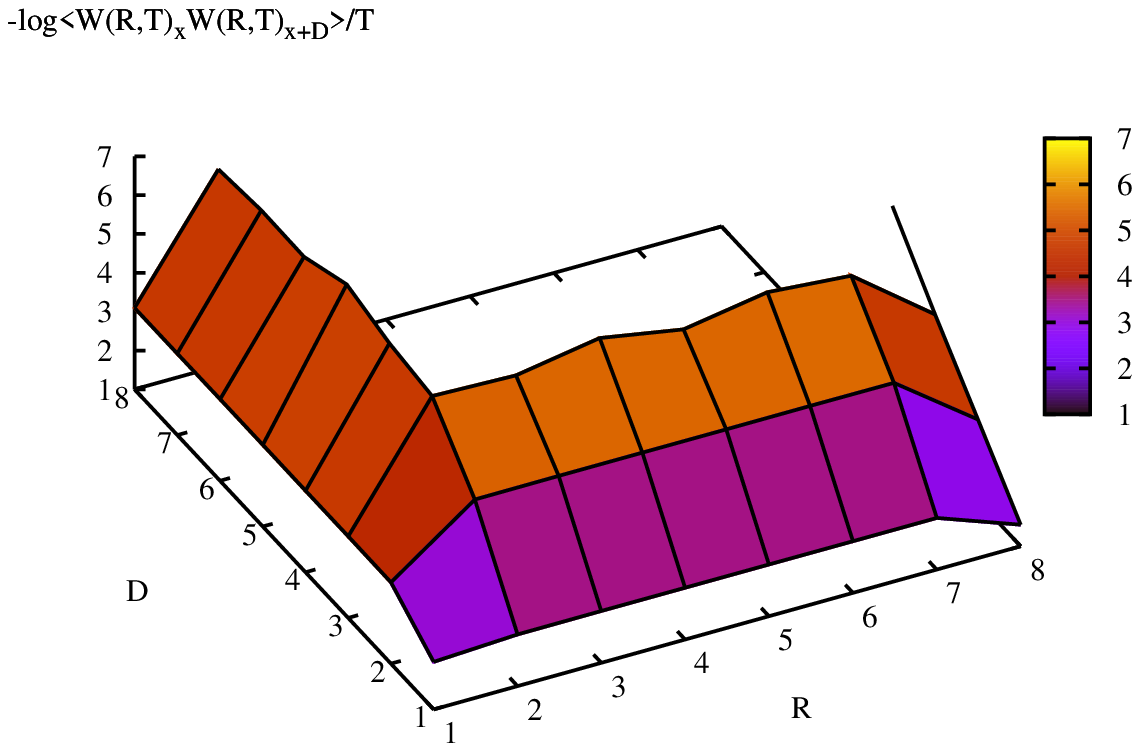}$\;$
	b)\includegraphics[width=.4\linewidth]{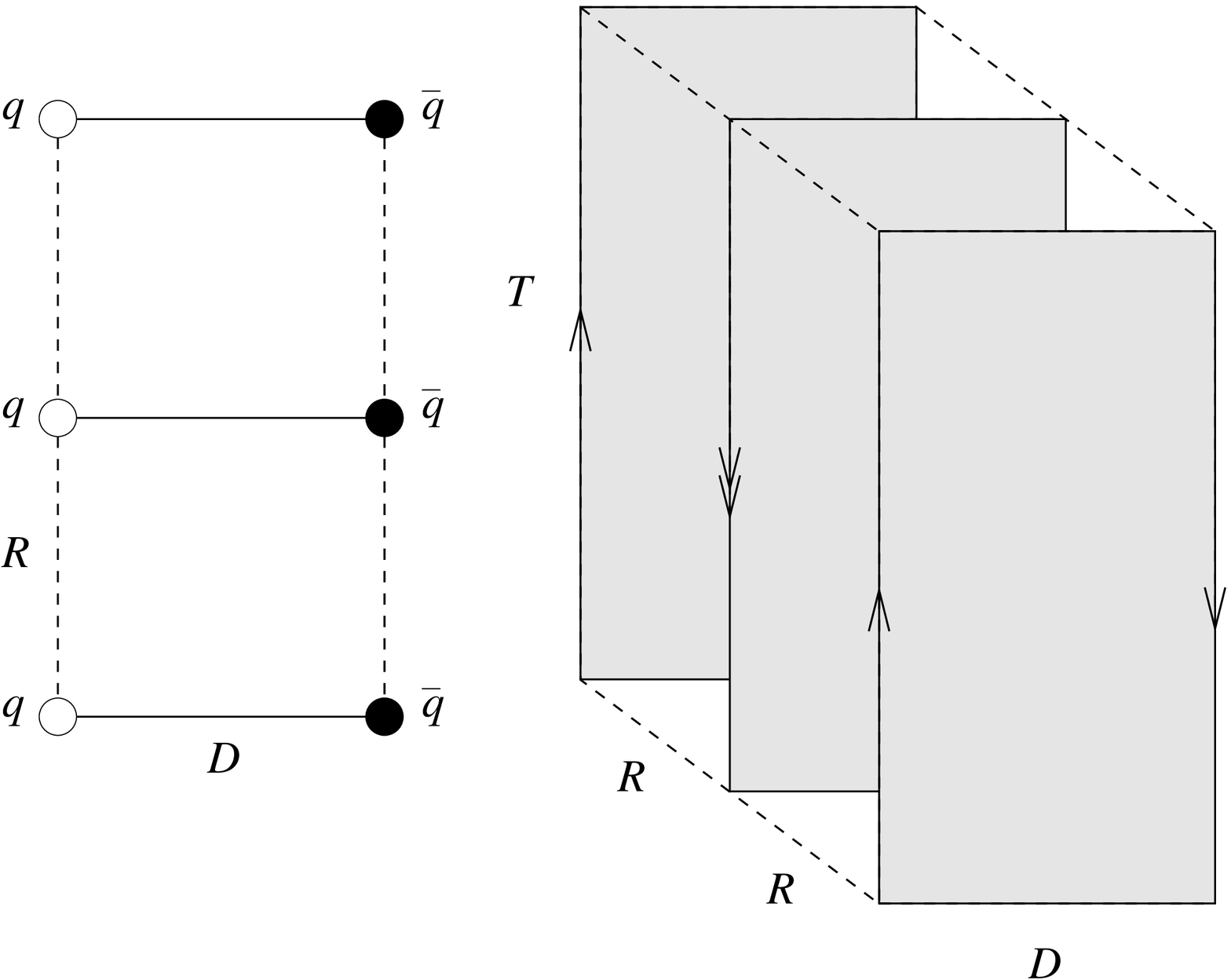}
	\caption{a) $Z(3)$ static baryon--anti-baryon potentials (Polyakov loop
correlators) $-log\langle W(R,T)_{x}W(R,T)_{x+D}\rangle/T$ for various
distances $R$, with $T=4$, between quarks (or anti-quarks), and distances
$D$ between static baryon and anti-baryon on $16^3\times4$ lattices. b) Polyakov
loop setup reflects the plateau ratios $4R\sigma:3D\sigma$ in
Fig.~\ref{fig:wcb}, {\it i.e.}, $4$ meson loops in $R$ direction vs.~$3$
in $D$ direction, meaning that, if $3D<4R$, bond rearrangement between the
quark sources changes the static baryon--anti-baryon into three static mesons.}
\label{fig:w3b}
\end{figure}

In Figs.~\ref{fig:wcb} and~\ref{fig:w3b}a we present 2D and 3D plots of static
baryon--anti-baryon potentials in the $Z(3)$ vortex model with respect to quark
source separations $R$ and baryon distances $D$. There is no more $R-D$
symmetry, which becomes clear in Fig.~\ref{fig:w3b}b, when after bond
rearrangement we do not correlate a static baryon and anti-baryon anymore, but
rather three static mesons. Hence, we have four mesonic loops making up the
static baryon and anti-baryon vs.~three static meson loops,
and as the same minimal area law behavior as above applies, the plateau
ratios $4R\sigma:3D\sigma$ in Fig.~\ref{fig:wcb} can be easily understood.
A nice detail is observed for static baryons with $R=8$
on our lattices with spatial lattice extent $16$: In this case, the outer
sources in the baryon and in the anti-baryon, respectively, share the same
spatial position and couple ($3\otimes3\rightarrow \bar{3} $,
$\bar{3} \otimes \bar{3} \rightarrow 3$), leaving a static meson-meson
correlator, as is evident from the constant $D$ data (green lines, dashed)
which drop at $R=8$ from $3D\sigma$ to $2D\sigma$. Thus, the potentials
again follow nicely the minimal area predictions; however, the signal
becomes limited owing to machine precision limitations for $R$ resp. $D>2$. 
Nevertheless, the potential energy redefined to asymptotically go to zero, 
{\it i.e.}, shifted by the asymptotic value $4R\sigma$ (two baryonic areas),
is again in good agreement with $SU(3)$ lattice studies~\cite{Inoue:2009ce},
keeping in mind the fact that center vortices do not reproduce a Coulomb
potential and
short distance effects can not be studied within the vortex models at fixed
lattice spacing $a=0.39$ fm. 

\clearpage

\noindent
{\small and in magenta (dashed) for constant $R$ data (red, solid).
%iffalse
	The plateau ratios $4R\sigma:3D\sigma$ can be understood from the Polyakov
	loop setup shown in Fig.~\ref{fig:w3b}b, it is simply the area ratio of the
	Polyakov loop correlators in $R$ vs. $D$ direction. 
%fi
	For constant $D$ plots (green, dashed), the data drop from the
$3D\sigma$ plateau to $2D\sigma$ at $R=8$ since, in this case, one is left
with a static meson-meson correlator because the two outer quarks (anti-quarks)
of the initial static baryon (anti-baryon) recombine to a anti-quark (quark)
source. The $R=8$ (red, solid) line in the bottom right plot (h) corresponds
to the static meson-meson correlator in Fig.~\ref{fig:mz3}h. Range of extracted
lattice data is limited by machine precision.}

\section{Conclusions}\label{sec:con}

Complementing earlier results within random center vortex world-surface models, 
which successfully describe the confinement-deconfinement transition and
reproduce (spatial as well as temporal) string tensions, topological
susceptibility and the (quenched) chiral condensate, we analyze static
meson-meson and baryon--anti-baryon potentials in $Z(2)$ and $Z(3)$ center
vortex backgrounds. To this end, we calculate multiple Polyakov loop
correlators corresponding to static meson-meson and baryon--anti-baryon
configurations in a hypercubic lattice model of random vortex world-surfaces,
an effective model for the infrared sector of Yang-Mills theory. An analysis
of this type, namely, of the area law associated with pairs of flat Wilson
loops in a center vortex background, similar to the analysis of multiple
Polyakov loop correlators performed here, had also been suggested previously
in \cite{Cornwall:2009as}.

We find that the expectation values of the static meson-meson and
baryon--anti-baryon correlators follow a minimal area law. Bond
rearrangement results when the mesons resp.~the baryon and anti-baryon become
too close to one another, {\it i.e.}, it becomes energetically favorable to
recombine the quark sources to two resp.~three complementary static meson
pairs. No evidence of long-distance tails in the
potentials, {\it i.e.}, van der Waals type forces, is found as the bonded
clusters are separated; the confining bonds in the clusters thus appear
to be truly saturating. Cluster separation occurs already at finite
distances, not only in an asymptotic sense.
These properties are expected for the confining dynamics governing quarks
in the strong interaction \cite{Lenz:1985jk}, and our results are in
good agreement with $SU(2)$ and $SU(3)$ lattice
studies~\cite{Fiebig:2005sw,Bali:2010xa,Inoue:2009ce} at long distances,
keeping in mind the fact that center vortices do not reproduce a Coulomb
potential, and short distance effects cannot be studied within the vortex
models at fixed lattice spacing $a=0.39$ fm.
It should be emphasized that there is no conclusive a priori argument in
favor of the (plausible) conjecture that the random vortex world-surface
model exhibits this behavior, {\it i.e.}, that it generates a strict
minimal area law at finite separations as described above. Thus, our study
constitutes a nontrivial test of the random vortex world-surface model,
investigating aspects of the confining dynamics of the model which had not
been previously probed. The model indeed reproduces the confinement
characteristics expected within the strong interaction.

Work is in progress to investigate static meson-meson correlators in random
center vortex models using quadratic and circular Wilson (not Polyakov) loop
correlators, focusing on the minimal area law~\cite{Hollwieser:2015qea}.
Motivated by the minimal surface of revolution problem, we aim to analyze
possible signs of catenary solutions.

\acknowledgments{We would like to thank Jarrett Moon for his collaboration in
the early stages of this work. This research was supported by the U.S. DOE
through the grant
DE-FG02-96ER40965 (D.A.,M.E.) and the Erwin Schr\"odinger Fellowship program of
the Austrian Science Fund FWF (``Fonds zur F\"orderung der wissenschaftlichen
Forschung'') under Contract No. J3425-N27 (R.H.). Calculations were performed on
the Phoenix and Vienna Scientific Clusters (VSC-2 and VSC-3) at the Vienna
University of Technology and the Riddler Cluster at New Mexico State University.}

\bibliographystyle{utphys}
\bibliography{../literatur}

\end{document}